\documentclass[aps,prb,twocolumn,
			   groupedaddress,superscriptaddress,
			   amsfonts,amssymb,amsmath, 
			   citeautoscript,
			   a4paper
			   ]{revtex4-1}

\usepackage{microtype} 

\usepackage{txfonts}  
\usepackage{txfontsb} 
\usepackage{bm} 

\usepackage{xcolor}
\usepackage{graphicx}

\usepackage{hyperref}
\hypersetup{
    colorlinks,
    linkcolor={blue!75!black!80!yellow},
    citecolor={blue!75!black!80!yellow},
    urlcolor={blue!75!black!80!yellow}
}

\usepackage[centering,hmargin=1.85cm,tmargin=31mm,bmargin=37mm]{geometry}

\newcommand{\rv}{\mathbf{r}}
\newcommand{\Ev}{\mathbf{E}}
\newcommand{\ef}{\epsilon_{\text{\textsc{f}}}}
\newcommand{\vf}{v_{\text{\textsc{f}}}}
\newcommand{\tsc}[1]{\text{\textsc{#1}}}
\newcommand{\e}{\mathrm{e}}
\newcommand{\dd}[1]{\mathrm{d}#1\,}
\newcommand{\appropto}{\mathrel{\vcenter{
  \offinterlineskip\halign{\hfil$##$\cr
    \propto\cr\noalign{\kern.2pt}\sim\cr\noalign{\kern-2.5pt}}}}}
    
\makeatletter
\renewcommand{\fnum@figure}{\figurename~\thefigure\ (color online)}
\makeatother

\begin{document}
\title{Localized plasmons in graphene-coated nanospheres}

\author{Thomas~Christensen}
\affiliation{Department of Photonics Engineering, Technical University of Denmark, DK-2800 Kgs. Lyngby, Denmark}
\affiliation{Center for Nanostructured Graphene, Technical University of Denmark, DK-2800 Kgs. Lyngby, Denmark}
\author{Antti-Pekka~Jauho}
\affiliation{Center for Nanostructured Graphene, Technical University of Denmark, DK-2800 Kgs. Lyngby, Denmark}
\affiliation{Department of Micro- and Nanotechnology, Technical University of Denmark, DK-2800 Kgs. Lyngby, Denmark}
\author{Martijn~Wubs}
\affiliation{Department of Photonics Engineering, Technical University of Denmark, DK-2800 Kgs. Lyngby, Denmark}
\affiliation{Center for Nanostructured Graphene, Technical University of Denmark, DK-2800 Kgs. Lyngby, Denmark}
\author{N.~Asger~Mortensen}
\email{asger@mailaps.org}
\affiliation{Department of Photonics Engineering, Technical University of Denmark, DK-2800 Kgs. Lyngby, Denmark}
\affiliation{Center for Nanostructured Graphene, Technical University of Denmark, DK-2800 Kgs. Lyngby, Denmark}


\begin{abstract}
We present an analytical derivation of the electromagnetic response of a spherical object coated by a conductive film, here exemplified by a graphene coating. Applying the framework of Mie--Lorenz theory augmented to account for a conductive boundary condition, we derive the multipole scattering coefficients, modified essentially through the inclusion of an additive correction in numerator and denominator. By reductionist means, starting from the retarded response, we offer simple results in the quasistatic regime by analyzing the multipolar polarizability and associated dispersion equation for the localized plasmons. We consider graphene coatings of both dielectric and conducting spheres, where in the former case the graphene coating introduces the plasmons and in the latter case modifies in interesting ways the existing ones. Finally, we discuss our analytical results in the context of extinction cross-section and local density of states. Recent demonstrations of fabricated spherical graphene nanostructures make our study directly relevant to experiments.
\end{abstract}

\maketitle

\section{Introduction}
The study of interaction between electromagnetic fields and graphene has seen a riveting development in recent years. In particular, efforts have centered on oscillation-energies near the experimentally achievable Fermi level of graphene, typically in the sub-eV range, where the principal features are due to the excitation of either propagating or localized two-dimensional plasmons\cite{Grigorenko:2012,Bludov:2013,Abajo:2014}. A large variety of structural configurations has been investigated, ranging from e.g., \mbox{(semi-)}finite structures\cite{Wang:2011,Chen:2012,Fei:2012,Thongrattanasiri:2013a} to periodic arrays\cite{Thongrattanasiri:2012a,Fang:2012,XZhu:2014}. Of these studies, the overwhelming majority exhibit the common assumption of structural planarity. Recently, interest has emerged also in exploring the properties of plasmonic-interaction in curved configurations, e.g., propagating modes in bent and corrugated sheets\cite{BingLu:2013}, in the context of cloaking\cite{ChenAlu:2012,ChenAlu:2013}, and in various coated nanowire systems\cite{Zhu:2014,Gao:2014a,Gao:2014b,Huang:2014}.

\begin{figure}[!htb]
\includegraphics[scale=.9]{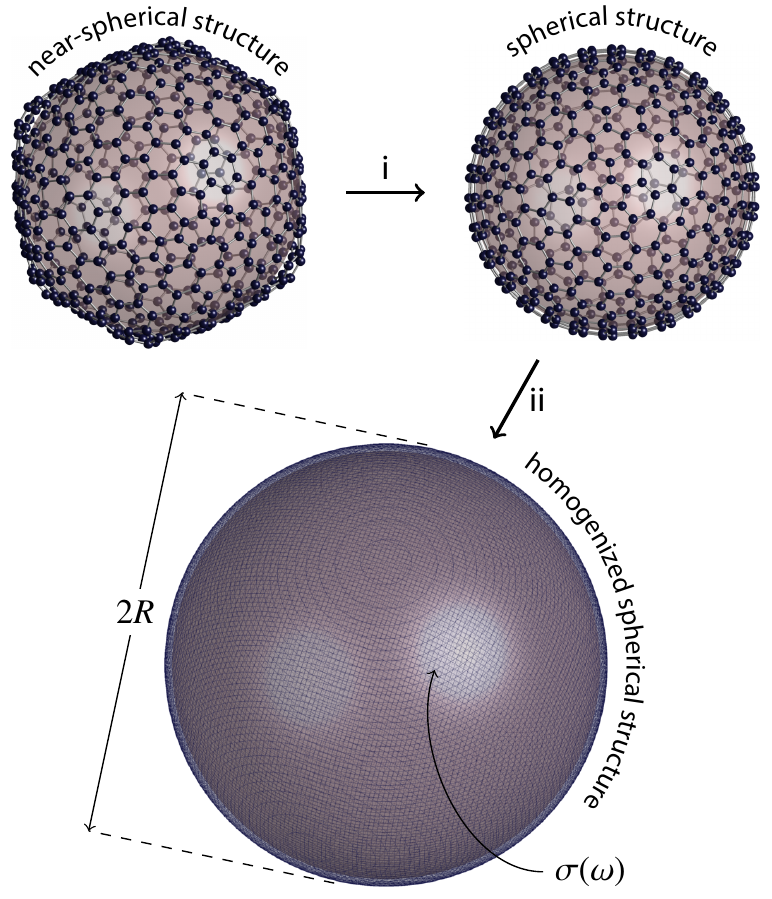}
\caption{Schematic illustration of the introduced conceptual simplifications in the treatment of optical response of graphene nanospheres via a surface-conductivity approach. Specifically at step \textsf{i)} any aspherical elements are neglected, while, at step \textsf{ii)} the microscopic details of the structure are replaced by the bulk response function $\sigma(\omega)$. Here depicted for a $C_{540}$ fullerene, for illustrative purposes solely.}\label{fig:setup}
\end{figure}

In this paper, we examine the archetypal curved graphene structure: a sphere, highly amenable to analytics and exhibiting the key features necessary to gain clear physical insight in the role of curvature. The spherical geometry is also of experimental relevance, given recent fabricational demonstrations. Notably, demonstrations include reduced graphene-oxide hollow spheres\cite{Yang:2014}, graphene-encapsulation of hollow SnO$_2$ spheres of radii down to $\sim\! 50\text{ nm}$\cite{Yang:2014}, and porous multilayer graphene nanospheres supported by a polystyrene interior\cite{Lee:2013}. Though the graphene in these recent demonstrations exhibits several deviations from an idealized two-dimensional spherical coating, they underscore the relevance of the geometry beyond a theoretical perspective. 
At the opposite end of the size-spectrum, the fullerenes represent a tempting analogy. However, it is now well-established that larger fullerenes, such as C$_{320}$ and beyond, prefer faceted, predominately icosahedral rather than spherical configurations\cite{Bakowies:1995,Calaminici:2009}. Additionally, the use of semiclassical, local response functions in graphene is reasonable only for structures in excess of $\sim\! 10^4$ carbon atoms\cite{Thongrattanasiri:2012}. In this paper, we take a classical, local surface conductivity approach, and as such we expect our predictions to be reliable chiefly for radii larger than $\sim\!5\text{ nm}$ (a graphene-sphere of $N$ atoms has a radius of $R\sim\! \sqrt{N}\times 0.457\text{ \AA}$), significantly beyond the range of fullerenes.

The paper is structured as follows: For graphene described by a local surface conductivity, we show in Section~\ref{sec:vectorwaves} that its electromagnetic response can be understood through a modified Mie--Lorenz theory. From the asymptotic limit of these results to the quasistatic realm, we derive the multipolar polarizability in Section~\ref{sec:nonretarded}, and identify the multipole plasmon conditions in the quasistatic regime, which are particularly transparent. Specifically, we show that an intuitive, effective-momentum mapping connects the sphere-resonances with those of an infinite plane. In Section~\ref{sec:dielectric} we present calculations for graphene-coated dielectric spheres, considering first the size-dispersion of the multipole resonances. Next, comparing two observables, the extinction cross-section and the local density of states (LDOS), we highlight the physical significance of the dipole and higher-order multipoles. In Section~\ref{sec:drude} we study the interaction between a localized plasmon supported by a Drude sphere, e.g., a doped semiconductor, and the plasmon supported by the graphene coating. We close our treatment of coated Drude spheres by discussing a corollary of the formalism related to surficial damping in metal plasmonics. Finally, we summarize the results in Section~\ref{sec:summary}.

\section{Theoretical Description}
Within the local-response approximation (LRA), the electric field $\Ev(\rv,\omega)$ in a homogeneous region $\mathcal{V}_{\!j}$ with dielectric constant $\varepsilon_{\!j}(\omega)$ satisfies the homogeneous Helmholtz wave equation
\begin{equation}\label{eq:helmholtz}
\nabla^2\Ev(\rv,\omega) - k_0^2\varepsilon_{\!j}(\omega)\Ev(\rv,\omega) = 0,
\end{equation}
where $k_0=\omega/c$ denotes the free-space wave number and where $\varepsilon_{\!j}(\omega)$ denotes the effective LRA dielectric constant, potentially exhibiting a frequency dependence.

\subsection{Retarded Solution by Expansion in Vector Waves}\label{sec:vectorwaves}
We solve Eq.~\eqref{eq:helmholtz} by expansion in vector wave functions, as befit for structures with curvilinear symmetries~\cite{Stratton}. In particular, within the LRA the electric field is divergence-free, or solenoidal, in which case the monochromatic solutions of the electric field in a homogeneous region $\mathcal{V}_j$ can be expanded in the basis of the solenoidal vector wave functions $\mathbf{M}_{\nu}^{[i]}(\rv)$ and $\mathbf{N}_{\nu}^{[i]}(\rv)$ of that region
\begin{equation}
\boldsymbol{\mathrm{E}}(\rv) =
\sum_{i\nu} 
 a_{\nu}\mathbf{M}_{\nu}^{[i]}(\rv)
+b_{\nu}\mathbf{N}_{\nu}^{[i]}(\rv),
\end{equation}
with $\nu$ denoting a geometry-dependent expansion index, while $i$ denotes expansion over in- and outgoing waves, and, finally, with $a_{\nu}^{[i]}$ and $b_{\nu}^{[i]}$ denoting associated expansion coefficients. The functions $\mathbf{M}_{\nu}^{[i]}(\rv)$ and $\mathbf{N}_{\nu}^{[i]}(\rv)$ describe the TE and TM parts, respectively, of the electric field, and describe the propagation of transverse modes, cf.\ their solenoidal property. In spherical coordinates $\rv=(r,\theta,\varphi)$ the index $\nu$ partitions into polar and azimuthal quantum numbers, $l\in[1,\infty[$ and $m\in[-l,l]$, with associated vector waves (usually referred to as multipoles)\cite{Stratton}
\begin{subequations}
\begin{align}
\mathbf{M}_{lm}^{[i]}(\rv) &= \nabla\times\rv\psi_{lm}^{[i]}(\rv),\\
\mathbf{N}_{lm}^{[i]}(\rv) &= \frac{1}{k} \nabla\times\nabla\times\rv\psi_{lm}^{[i]}(\rv),
\end{align}
\end{subequations}
defined in terms of the scalar generating functions $\psi_{lm}^{[i]}(r,\theta,\varphi) = z_l^{[i]}(kr)P_l^m(\cos\theta)\e^{im\varphi}$, where $z_l^{[i]}$ denotes spherical Bessel or Hankel functions (of the first kind), $j_l$ and $h_l^{\scriptscriptstyle(1)}$, for $i=1$ and $i=2$, respectively, corresponding to in- and outgoing waves.
The wave number $k\equiv k_0\!\sqrt{\varepsilon_j}$ relates the dimensionless argument $kr$ with the material properties.
An additional class of vector wave functions exists, denoted $\mathbf{L}_{\nu}(\rv)$, which are irrotational. These vector waves are needed e.g., in the description of longitudinal modes arising in nonlocal response or in the presence of sources, but are irrelevant in homogeneous media described by the LRA\cite{Ruppin:1973a,David:2012,Christensen:2014}.

Here we consider the specific case of a two-component spherically symmetric system, centered at origo, coated by a conductive film at the bulk-component interface at radius $R$, as indicated in Fig.~\ref{fig:setup}. We assume that the system is illuminated from a source in the external region, denoted $\mathcal{V}_2$, by the ingoing field $\Ev^{\mathrm{inc}}(\rv)$. The incident field excites an outgoing scattered field, $\Ev^{\mathrm{sca}}(\rv)$, in $\mathcal{V}_2$, and an ingoing transmitted field, $\Ev^{\mathrm{tra}}(\rv)$, in the interior region, denoted $\mathcal{V}_1$. Explicitly, the field in- and outside the sphere is expanded via
\begin{subequations}
\begin{align}
\Ev_{\scriptscriptstyle\mathcal{V}_1}(\rv) =&{}
\sum_{lm} a_{lm}^{\mathrm{tra}} \mathbf{M}_{lm}^{[1]}(\rv) + b_{lm}^{\mathrm{tra}}\mathbf{N}_{lm}^{[1]}(\rv),\space\space\space &r<R,\\ \nonumber
\Ev_{\scriptscriptstyle\mathcal{V}_2}(\rv) =&{} 
\sum_{lm} a_{lm}^{\mathrm{inc}} \mathbf{M}_{lm}^{[1]}(\rv) + b_{lm}^{\mathrm{inc}}\mathbf{N}_{lm}^{[1]}(\rv)\\ 
&+\sum_{lm}a_{lm}^{\mathrm{sca}} \mathbf{M}_{lm}^{[2]}(\rv) + b_{lm}^{\mathrm{sca}}\mathbf{N}_{lm}^{[2]}(\rv),\space\space\space &r>R,
\end{align}
\end{subequations}
where regions $\mathcal{V}_{\!j}$ are implicitly associated with wave numbers $k_{\!j}=k_0\!\sqrt{\varepsilon_{\!j}}$. 

The transmitted and scattered amplitudes, $\{a_{lm}^{\mathrm{tra}},b_{lm}^{\mathrm{tra}}\}$ and $\{a_{lm}^{\mathrm{sca}},b_{lm}^{\mathrm{sca}}\}$, are linearly proportional to the incident amplitudes, $\{a_{lm}^{\mathrm{inc}},b_{lm}^{\mathrm{inc}}\}$. Their interrelation is dictated by the boundary conditions (BCs) at the domain-interface at $r=R$. In the presence of a conductive surface at $r=R$ a finite surface current $\mathbf{K}$ is introduced, in which case the BCs read as $\hat{\mathbf{n}}\times(\Ev_{\scriptscriptstyle \mathcal{V}_2}-\Ev_{\scriptscriptstyle \mathcal{V}_1}) = 0$ and $\hat{\mathbf{n}}\times(\mathbf{H}_{\scriptscriptstyle \mathcal{V}_2}-\mathbf{H}_{\scriptscriptstyle \mathcal{V}_1}) = \mathbf{K}$ at all surficial points. We take the induced surface current at a surficial point $\rv$, with associated normal $\hat{\mathbf{n}}$, as linearly related to the tangential field $\Ev_{\scriptscriptstyle\parallel}(\rv)$, constructed such that $\Ev_{\scriptscriptstyle\parallel}(\rv)\cdot \hat{\mathbf{n}} = 0$, via an LRA surface conductivity $\sigma(\omega)$, such that $\mathbf{K}(\rv) = \sigma(\omega)\Ev_{\scriptscriptstyle\parallel}(\rv)$.

Enforcing these BCs translates into local, linear relations between the scattered and incident amplitudes
\begin{equation}
a_{lm}^{\mathrm{sca}} = t_{l'}^{\text{\textsc{te}}}a_{l'm'}^{\mathrm{inc}} \delta_{ll'}\delta_{mm'},\qquad
b_{lm}^{\mathrm{sca}} = t_{l'}^{\text{\textsc{tm}}}b_{l'm'}^{\mathrm{inc}} \delta_{ll'}\delta_{mm'},
\end{equation}
where the proportionality constants, often referred to as Mie--Lorenz scattering coefficients, are given by
\begin{widetext}
\begin{subequations}\label{eqs:Miecoefs}
\begin{align}
t_l^{\text{\textsc{te}}} &= 
\frac{ -j_l(x_1)[x_2 j_l(x_2)]' + j_l(x_2)[x_1 j_l(x_1)]' - g(\omega) x_0^2 j_l(x_1) j_l(x_2)}
{j_l(x_1)[x_2 h_l^{\scriptscriptstyle (1)}(x_2)]' - h_l^{\scriptscriptstyle (1)}(x_2)[x_1 j_l(x_1)]' + g(\omega) x_0^2 j_l(x_1) h_l^{\scriptscriptstyle (1)}(x_2)},\\
t_l^{\text{\textsc{tm}}} &= 
\frac{ -x_2^2 j_l(x_2) [x_1 j_l(x_1)]' + x_1^2 j_l(x_1) [x_2 j_l(x_2)]' + g(\omega)x_0^2 [x_1 j_l(x_1)]' [x_2 j_l(x_2)]'}
{x_2^2 h_l^{\scriptscriptstyle (1)}(x_2) [x_1 j_l(x_1)]' - x_1^2 j_l(x_1) [x_2 h_l^{\scriptscriptstyle (1)}(x_2)]' - g(\omega)x_0^2 [x_1 j_l(x_1)]' [x_2 h_l^{\scriptscriptstyle (1)}(x_2)]'},
\end{align}
\end{subequations}\vspace*{-1em}
\end{widetext}
written in terms of the dimensionless argument $x_j \equiv k_j R$ for $j=\{0,1,2\}$, and where the influence of the conductive surface is included via the dimensionless parameter
\begin{equation}
g(\omega) \equiv \frac{i\sigma(\omega)}{\varepsilon_0\omega R}.
\end{equation}
Naturally, for vanishing surface conductivity $g(\omega)\rightarrow 0$, whereby the solution reduces to the standard Mie--Lorenz coefficients\cite{BohrenHuffman:1983}.

\subsection{The Multipolar Polarizability and the Nonretarded Plasmon Dispersion}\label{sec:nonretarded}
The amplitudes in Eq.~\eqref{eqs:Miecoefs} give the fully retarded response. However, as is well known, the quasistatic limit is excellent in the context of plasmonic excitations in graphene when $\hbar\omega/\ef\gg \alpha\approx 1/137$\cite{Koppens:2011}. For optical interactions in the quasistatic regime, the multipolar polarizability, $\alpha_l$, constitutes the central object, and can be derived from the TM Mie--Lorenz coefficients via\cite{David:2012}
\begin{equation}\label{eq:MieLorenzcoefs}
\alpha_l = -4\pi i\frac{l[(2l+1)!!]^2}{(l+1)(2l+1)}\lim_{x_0\rightarrow 0}\Bigg[\frac{t_l^{\text{\textsc{tm}}}}{k_2^{2l+1}}    \Bigg],
\end{equation}
with $!!$ denoting the double factorial. From this we can derive (using the small-argument asymptotic expansions of the spherical Bessel functions) the multipolar polarizability in the quasistatic limit
\begin{equation}\label{eq:alpha_l}
\alpha_l = 4\pi R^{2l+1} \frac{l[\varepsilon_1-\varepsilon_2+(l+1)g(\omega)]}{l\varepsilon_1+(l+1)\varepsilon_2+l(l+1)g(\omega)}.
\end{equation}
This expression is naturally highly reminiscent of the well-known result for the polarizability of a two-component spherical system\cite{Fuchs:1987,Myroshnychenko:2008}, but extended via $g(\omega)$ to account for the presence of a conductive surface. 

The plasmonic resonances of the system are obtained from the poles of the Mie--Lorenz coefficients of Eq.~\eqref{eq:MieLorenzcoefs}, or, in the quasistatic regime, from the poles of the polarizability of Eq.~\eqref{eq:alpha_l}. In the latter case, we can derive a uncomplicated resonance condition for the $l$-order multipolar plasmon at frequency $\omega_l$, extending the Fr\"{o}hlich condition to account for a conductive surface contribution
\begin{equation}\label{eq:disprel_stat}
l\varepsilon_1 + (l+1)\varepsilon_2 + l(l+1)g(\omega_l) = 0.
\end{equation}
Though usually -- in the absence of a coating -- the existence of a plasmon requires $\varepsilon_1\varepsilon_2<0$, it is evident that plasmons may exist here even when $\varepsilon_1,\varepsilon_2>0$ provided that $g(\omega_l)$ is sufficiently negative, achievable for a surface conductivity with $\text{Im}(\sigma)<0$.

\begin{figure}[!b]\centering
\includegraphics[scale=1]{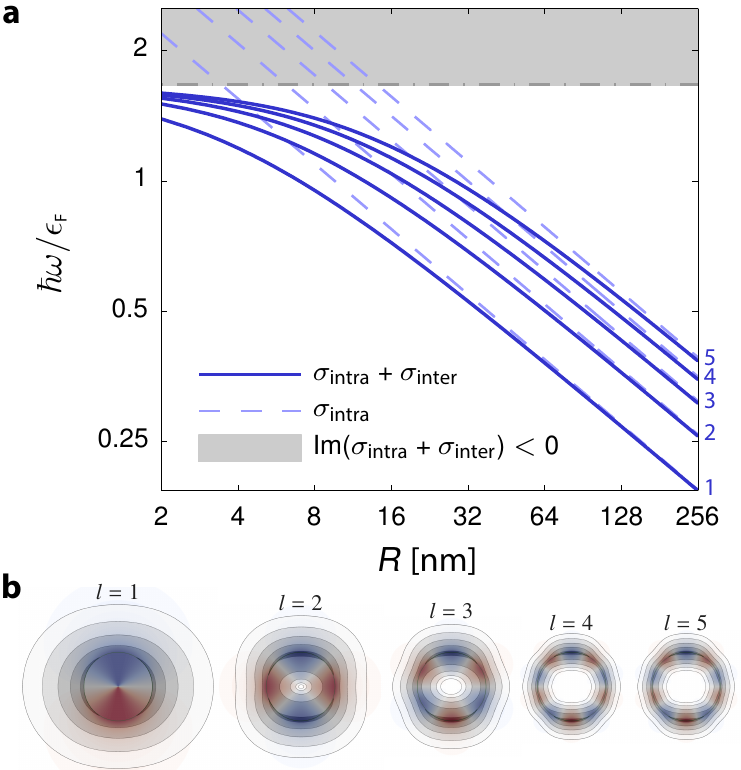}
\caption{\textsf{(\textbf{a})} Dispersion of the plasmon resonance-frequency as a function of sphere-radius for the first five multipole-plasmons ($l$ indicated in blue text) depicted in a doubly-logarithmic plot. Resonances calculated for lossless graphene spheres with Fermi level $\ef = 0.4\ \mathrm{eV}$ in vacuum ($\varepsilon_1=\varepsilon_2=1$) at zero temperature. The intra-band approximation (discussed in the text) is indicated in dashed blue lines, while the region of negative imaginary part of the conductivity, $\hbar\omega/\ef\gtrsim 1.6671$, is indicated in gray shading. \textsf{(\textbf{b})} Induced multipole modal profiles in the $xz$ plane, calculated for $R=20\text{ nm}$ and $m=1$ at resonance. Indicated is $|\mathbf{E}|$ in contours (separated by factors of 2), and $\mathrm{Re}(E_\theta)$ in blue and red, corresponding to positive and negative, respectively.}\label{fig:dispersion}
\end{figure}

For a uniform background, $\varepsilon_1=\varepsilon_2=\varepsilon$, the condition is particularly elucidating, reading
\begin{equation}
\frac{2i\varepsilon_0\varepsilon\omega_l}{\sigma(\omega_l)} = \bigg(1+\frac{1}{2l+1}\bigg)\frac{l}{R}\equiv q_l^{\mathrm{eff}},
\end{equation}
here we have cast the condition in the equivalent form as that of the infinite sheet plasmon condition,\cite{Low:2014} whereby we are able to identify an effective momentum $q_l^{\mathrm{eff}}$, which, rather suggestively, approaches $l/R$ asymptotically as $l\rightarrow\infty$, as a consequence of the modes perceiving the curving surface as increasingly flat with higher $l$ and concomitant shorter wavelengths\cite{Myroshnychenko:2008,Christensen:2014}. 
For the optically important dipole resonance, we find $q_1^{\mathrm{eff}} = \tfrac{4}{3}R^{-1}$.

The identification of an effective momentum suggests a phenomenological approach to incorporate the effects of nonlocal response (momentum dispersion), by substituting $\sigma(\omega)\rightarrow \sigma(q_l^{\mathrm{eff}},\omega)$, with the latter expression obtainable e.g., from the noninteracting polarizability\cite{Wunsch:2006,Hwang:2007}. However, though such a speculative approach certainly is alluring, it would constitute an overextension of the momentum analogy. Indeed, upon including nonlocal response through its hydrodynamic approximation one finds that the correct effective momentum takes a form $q_l^{\text{eff,\textsc{h}}}\equiv\sqrt{l(l+1)}/R$, clearly distinct from $q_l^{\mathrm{eff}}$. For completeness we discuss the inclusion of hydrodynamic response in the conductive coating in Appendix~\ref{app:hydrodynamic}, whose contribution can be accounted for by a straightforward rescaling of the local-response conductivity.

\begin{table}
\centering
\includegraphics{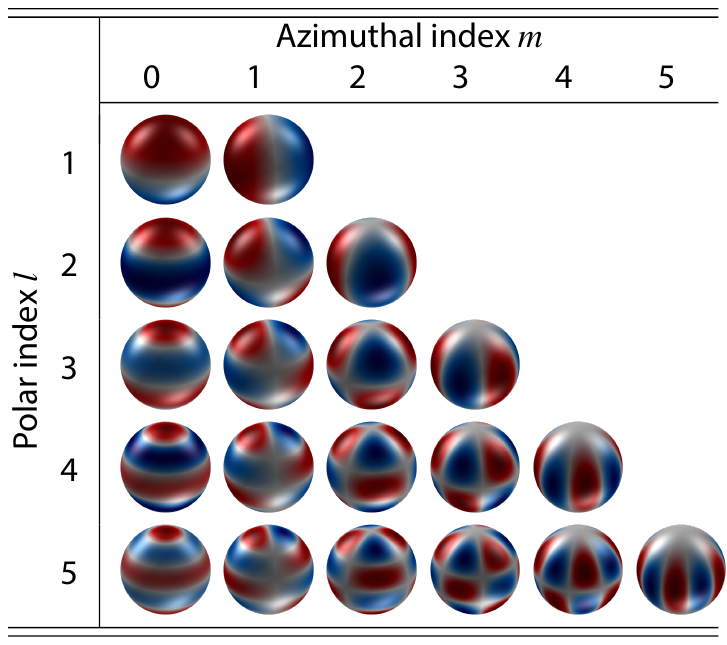}
\caption{Representation of the induced charge profiles of the $lm$-multipole plasmons. The charge profile is $\rho_{lm}(\theta,\varphi)\propto P_l^m(\cos \theta)\e^{im\varphi}$ of which we here depict the real part of the latter. Negative $m$-values differ from their positive counterparts only by rotational direction.}\label{tab:InducedCharge}
\end{table}

\section{Graphene coating of dielectric spheres}\label{sec:dielectric}
The analysis so far is valid for any spectral dependence of $\sigma(\omega)$ (or, indeed, of $\varepsilon_1$ or $\varepsilon_2$). For a graphene-coated system, we take $\sigma(\omega)$ as graphene's bulk LRA conductivity, which, for a Fermi level $\ef$ at finite temperature $T$ reads as $\sigma(\omega) = \sigma_{\text{intra}}(\omega) + \sigma_{\text{inter}}(\omega)$\cite{Falkovsky:2007}
\begin{subequations}
\begin{align}
\frac{\sigma_{\mathrm{intra}}(\omega)}{\sigma_0} &= 
\frac{2ik_{\tsc{b}}T}{\hbar\tilde{\omega}}
\ln\!\Bigg[2\cosh\Bigg(\frac{\ef}{2 k_{\tsc{b}}T}\Bigg)\Bigg],\\
\frac{\sigma_{\mathrm{inter}}(\omega)}{\sigma_0} &=
\frac{\pi}{4} H\big(\tfrac{1}{2}\hbar\omega\big) + i\hbar\tilde{\omega}\!
\int_0^\infty \!\!\!\! \dd{\epsilon}\frac{H(\epsilon) - H\big(\tfrac{1}{2}\hbar\omega\big) }
{\hbar^2\tilde{\omega}^2 - 4\epsilon^2},
\end{align}
with definitions $\tilde{\omega} \equiv \omega+i\gamma_{\text{g}}$ where $\gamma_{\text{g}}$ denotes the optical loss-rate of graphene\cite{Gusynin:2007,Hanson:2008}, $\sigma_0 \equiv e^2/\pi\hbar$ is the quantum of conductance, and $H(\epsilon)$ the population difference between energies $\mp \epsilon$
\begin{equation}
H(\epsilon) = \frac{\sinh(\epsilon/k_{\tsc{b}}T)}
{\cosh(\ef/k_{\tsc{b}}T)+\cosh(\epsilon/k_{\tsc{b}}T)}.
\end{equation}
\end{subequations}
In the ensuing subsections we consider nondispersive bulk media, i.e.,\ spectrally constant $\varepsilon_1$ and $\varepsilon_2$, that is, dielectrics. In this case, the existence of localized plasmons is strictly due to the graphene coating. In Section~\ref{sec:drude} we explore a dispersive interior, concretized by a graphene-coated Drude sphere, with the accompanying plasmons emerging from the  interaction of the plasmon-branches of each bare subsystem.

\subsection{Size dispersion and modal profile}
In Fig.~\ref{fig:dispersion}(a) we investigate the size dispersion of the plasmonic modes of graphene spheres in vacuum in the low-temperature, low-loss limit, by solving Eq.~\eqref{eq:disprel_stat} numerically. It is evident that for large spheres and sufficiently low $l$, the intraband, low-loss approximation, $\sigma(\omega)\simeq \sigma_{\text{intra}}(\omega)|_{T= 0}^{\gamma_{\text{g}}=0} = ie^2\ef/\pi\hbar^2\omega$, is a good approximation, yielding the dispersion $\hbar\omega_l\simeq \big[e^2\ef/\pi\varepsilon_0\varepsilon^{\text{\textsc{b}}}_l R\big]^{1/2}$ with $\varepsilon^{\text{\textsc{b}}}_l \equiv \varepsilon_1/(l+1) + \varepsilon_2/l$. For smaller spheres, and concomitant larger resonance frequencies, the interband term redshifts the resonances significantly\cite{Wang:2013}. Furthermore, since $\text{Im}(\sigma_{\text{intra}}+\sigma_{\text{inter}})$ changes sign from positive to negative at $\hbar\omega/\ef\approx 1.6671$ the LRA predicts plasmon resonances restricted to the range below this frequency -- though the inclusion of nonlocal response relaxes this restriction\cite{Wunsch:2006,Hwang:2007}. 
As seen from Fig.~\ref{fig:dispersion}(b) the electric fields associated with each multipole plasmon are increasingly confined to the surface region with increasing $l$, in analogy with the increasing confinement experienced by a surface plasmon polariton with increasing momentum. In line with this analogy, the plasmon modes also exhibit a monotonically decreasing circumferential wavelength, displaying $l$ nodal lines of $E_{\theta}$ in the $xz$ plane for the $l$th mode. In general, as shown in Table~\ref{tab:InducedCharge}, the induced charge exhibits exactly $l$ nodal lines, regardless of the value of $m$.

\begin{figure}[!htb]
\includegraphics[scale=1]{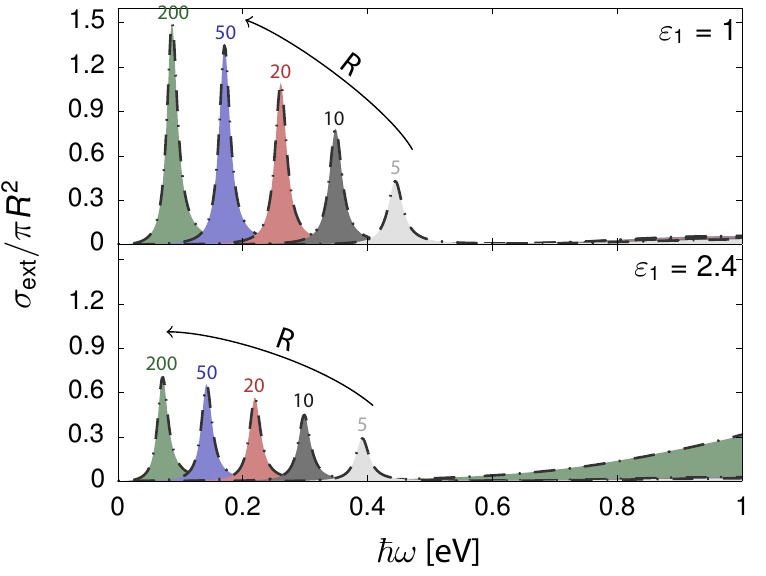}
\caption{Extinction cross-section for graphene spheres in vacuum of varying radii $R=5$, $10$, $20$, $50$, and $200$~nm (indicated in color), for loss-rate $\gamma_{\text{g}}=20\ \text{meV}$ and temperature $T=300\ \text{K}$. Top and bottom panels consider the interior spheres consisting of vacuum and polystyrene ($\varepsilon_1=1$ and $\varepsilon_1=2.4$), respectively. The quasistatic dipole approximation as well as fully retarded calculations are presented, here shown in dashed black lines and colored areas, respectively.}\label{fig:extinction}
\end{figure}

\subsection{Extinction and dipolar response}\label{sec:extinction}
For direct optical measurements the quantity of interest is typically the extinction cross-section, which is obtainable from either the Mie--Lorenz coefficients or, approximately, from the dipolar polarizability via\cite{BohrenHuffman:1983} $\sigma_{\mathrm{ext}} = 2\pi k_{2}^{-2}\sum_{l=1}^\infty(2l+1)\mathrm{Re}(t_l^{\text{\textsc{te}}}+t_l^{\text{\textsc{tm}}}) \simeq k_2\mathrm{Im}(\alpha_1)+(6\pi)^{-1}k_2^4|\alpha_1|^2$. As is evident from the quasistatic approximation of $\sigma_{\mathrm{ext}}$ only the dipole plasmon influences the cross-section in small spheres. In Fig.~\ref{fig:extinction} we show the extinction cross-sectional efficiency of graphene-coated spheres of vacuum and polystyrene, surrounded externally by vacuum. Indeed, it is evident that the dipole approximation is excellent, even for graphene-coated spheres of several hundreds of nanometers. As already observed in Fig.~\ref{fig:dispersion}(a), the resonance position is redshifted with increasing radius, leading to a size-dependent extinction cross-section. This stands in contrast to the resonances of metallic nanospheres which, in the classic quasistatic picture, exhibit size-independent resonances (though the inclusion of nonlocal response introduces a size-dependence\cite{Mortensen:2014}). The inclusion of a non-unity dielectric as the spherical substrate redshifts and lowers the overall response efficiency as seen from the calculations for coated polystyrene spheres ($\varepsilon_1=2.4$). This is consistent with the redshift generally arising from a reduction of the effective Coulomb interaction $1/\varepsilon^{\text{\textsc{b}}}_lR$ (since $\varepsilon^{\text{\textsc{b}}}_l$ increases with $\varepsilon_2$). 

\begin{figure}[!htb]
\includegraphics[scale=1]{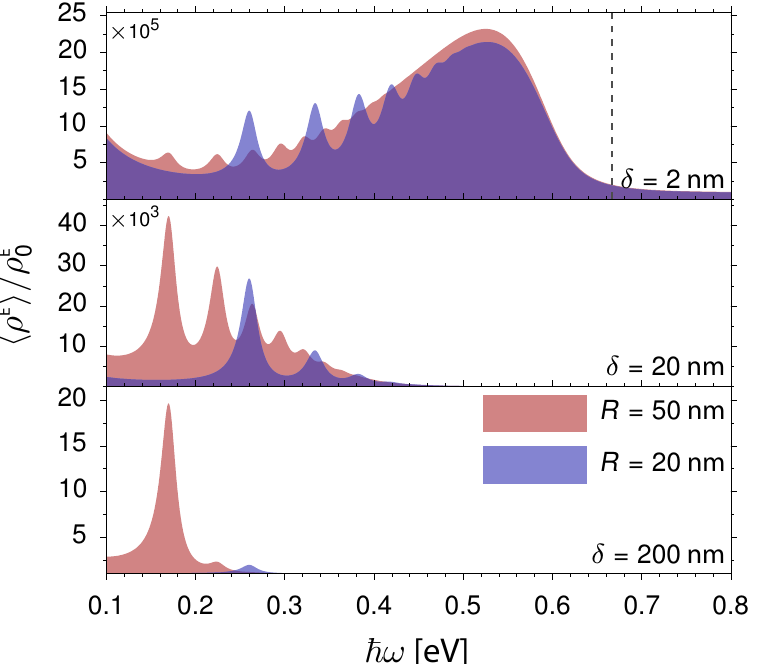}
\caption{LDOS-enhancement for graphene spheres in vacuum of varying radii $R=20$ and $50$~nm, with setup otherwise as in Fig.~\ref{fig:extinction}. The surface-to-observation distance $\delta$ is indicated in each panel. In the top panel, the $l\rightarrow\infty$ multipole asymptote at $\hbar\omega/\ef\approx 1.6671$ is indicated by the dashed line. For each panel, the $y$-axis ranges from unity and upwards.}\label{fig:ldos}
\end{figure}

\subsection{LDOS and multipolar response}
Exploring the properties of plasmons beyond the dipole resonance is best facilitated by near-field measurements whose excitation (and sampling profile) are not plane waves\cite{Christensen:2014}. Such non-planar exciting fields are naturally associated with nearby dipole emitters, such as dyes. The interplay between emitter and plasmonic system, leading e.g., to decay enhancement\cite{Koppens:2011}, is then governed by the electric LDOS.\cite{Joulain:2003} As for the cross-section, the LDOS-enhancement, i.e.,\ the LDOS near the nanosphere, $\rho^{\text{\textsc{e}}}$, relative to the LDOS in free-space, $\rho_0^{\text{\textsc{e}}}$, can be obtained from the Mie--Lorenz coefficients -- or, more simply, from the multipolar polarizability. In particular, in the quasistatic limit, the emitter-orientation-averaged LDOS-enhancement reads as\cite{Christensen:2014,Vielma:2007}
\begin{equation}
\frac{\langle\rho^{\text{\textsc{e}}}\rangle}{\rho^{\text{\textsc{e}}}_0} = 1 + \frac{1}{8\pi k_2^3}\sum_{l=1}^{\infty}(l+1)(2l+1)\frac{\mathrm{Im}(\alpha_l)}{(R+\delta)^{2(l+2)}},
\end{equation}
evaluated at radial distance $\delta$ from the sphere-surface.

In Fig.~\ref{fig:ldos} we consider the spectral dependence of the orientation averaged LDOS at varying distances from the spherical coating. At large surface-to-probe separations the LDOS is dominated by the dipolar plasmon, whilst, at increasingly shorter separations the high-order multipoles appear as significant features. The LDOS evaluated at just 2 nm above the surface reveals a broad multiple multipole feature. This broad feature is comprised of several spectrally overlapping multipole plasmons, that are increasingly congested as the energies approach $\approx 1.6671\ef$. Once more, this effect has a close analogue in metallic nanospheres where local response incurs a pile up of multipole modes near the flat-interface surface plasmon resonance at $\omega_{\mathrm{p}}/\sqrt{2}$, with $\omega_{\mathrm{p}}$ denoting the metallic plasma frequency.\cite{Myroshnychenko:2008, Christensen:2014}

\begin{figure*}[!htb]
\includegraphics[scale=1]{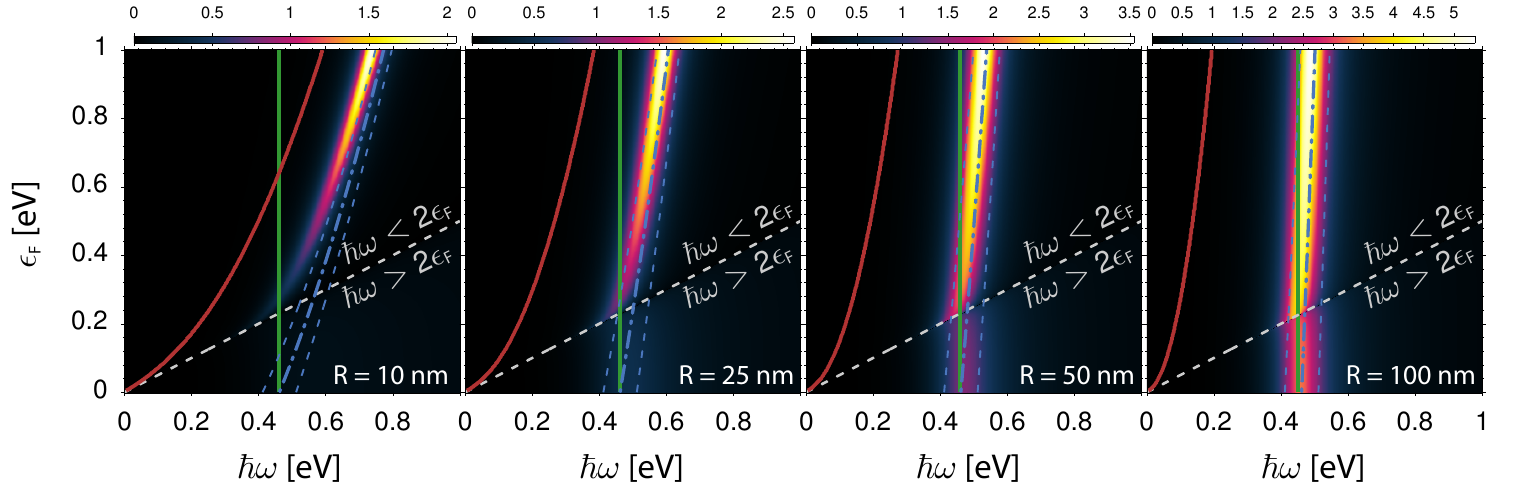}
\caption{Extinction cross-sectional efficiency in a graphene-coated Drude sphere in vacuum, explored as a function of graphene Fermi energy, frequency, for several fixed values of the radius (as indicated). The Drude material is characterized by its plasma frequency $\hbar\omega_{\mathrm{p}} = 0.8\ \text{eV}$ and loss-rate $\gamma_{\text{\textsc{d}}} = 0.1\ \text{eV}$, while graphene parameters are evaluated with $\gamma_{\text{g}}=20\ \text{meV}$ at $T=0\ \text{K}$. 
The color coded response indicates the extinction efficiency, $\sigma_{\text{ext}}/\pi R^2$, calculated from the fully retarded expression. The green and red lines indicate the retarded bare resonance positions of the Drude sphere and graphene-coated vacuum sphere, respectively. In blue is given the quasistatic intraband approximation, Eq.~\eqref{eq:drudegraphene_intra}, with the dash-dotted line indicating the resonance position, and the dashed lines the resonance width via $\omega_{\text{\textsc{r}}}\pm \omega_{\text{\textsc{i}}}$. Finally, the dashed white line separates the  regions of zero and nonzero Landau damping.\label{fig:drudeandgraphene}}
\end{figure*}

\section{Graphene coating of Drude spheres}\label{sec:drude}
Proceeding from the study of a non-dispersive interior, we consider next a graphene-coated Drude sphere, wherein we assign the interior dielectric function a Drude form
\begin{equation}\label{eq:drudeform}
\varepsilon_1(\omega) = \varepsilon_{\scriptscriptstyle\infty} - \frac{\omega_{\mathrm{p}}^2}{\omega(\omega+i\gamma_{\text{\textsc{d}}})},
\end{equation}
where $\varepsilon_{\scriptscriptstyle\infty}$ gives the residual high-frequency response of the material, and $\gamma_{\text{\textsc{d}}}$ the optical loss-rate of the Drude material. The Drude dispersion is traditionally applied to metals, but also reliably describes strongly doped semiconductors much larger in extent than the Fermi wavelength.\cite{Schimpf:2014,ZhangNordlander:2014} The case of doped semiconductor spheres is significantly more interesting from the perspective of mode hybridization, as the range of plasma frequencies of doped semiconductors overlap the realizable Fermi energies of graphene.

In this case where the interior is dispersive and well-described by Eq.~\eqref{eq:drudeform} the resonances of the coated system then follow directly from Eq.~\eqref{eq:disprel_stat}. 
If we include only the low-temperature intraband response of graphene, via $\sigma_{\text{intra}}(\omega) = ie^2\ef/\pi\hbar^2(\omega+i\gamma_{\text{g}})$, the dipole resonance condition is particularly simple, reading as
\begin{subequations}
\begin{equation}
\frac{\omega_{\mathrm{p}}^2}{\omega(\omega+i\gamma_{\text{\textsc{d}}})} + \frac{\omega_{\mathrm{g}R}^2}{\omega(\omega+i\gamma_{\text{g}})}  = \varepsilon_{\scriptscriptstyle\infty}+2\varepsilon_2,
\end{equation}
where 
\begin{equation}
\omega_{\mathrm{g}R}^2 \equiv \frac{2 e^2\ef}{\varepsilon_0\pi\hbar^2 R}
\end{equation}
\end{subequations}
plays the role of an effective plasma-frequency of the graphene coating. To first order in the loss-rates $(\gamma_{\text{\textsc{d}}},\gamma_{\text{g}})$ we find a single positive resonance frequency $\omega\equiv \omega_{\text{\textsc{r}}} - i\omega_{\text{\textsc{i}}}$ with\cite{Note1}
\begin{equation}\label{eq:drudegraphene_intra}
\omega_{\text{\textsc{r}}} \simeq \sqrt{\frac{\omega_{\mathrm{p}}^2 + \omega_{\mathrm{g}R}^2}{\varepsilon_{\scriptscriptstyle\infty}+2\varepsilon_2}},\qquad
\omega_{\text{\textsc{i}}} \simeq \frac{1}{2}\frac{\omega_{\mathrm{g}R}^2\gamma_{\text{g}} + \omega_{\mathrm{p}}^2\gamma_{\text{\textsc{d}}}}{\omega_{\mathrm{g}R}^2+\omega_{\mathrm{p}}^2}.
\end{equation}
The result bears a high resemblance with the standard quasistatic dipole resonance of a Drude sphere, but here lucidly adapted to account for the graphene coating through $\omega_{\mathrm{g}R}^2$. Considering $\omega_{\text{\textsc{i}}}$ it is interesting to note that for $\gamma_{\text{\textsc{d}}}\neq\gamma_{\mathrm{g}}$ it is possible to achieve an effectively reduced plasmon decay rate compared e.g., to the uncoated Drude sphere by appropriate scaling of $\omega_{\mathrm{g}R}$ relative to $\omega_{\mathrm{p}}$: specifically, if $\omega_{\mathrm{g}R}^2/\omega_{\mathrm{p}}^2\equiv a$ then $\omega_{\text{\textsc{i}}}\simeq \tfrac{1}{2(a+1)}(a\gamma_{\mathrm{g}}+\gamma_{\text{\textsc{d}}})$, illustrating that the decay can be tuned from predominately graphene- to Drude-like by varying the ratio $a$. Evidently, the comparative importance of graphene's response relative to the bulk Drude material's is indicated by the $a$, with the former dominant when $a>1$.

In Fig.~\ref{fig:drudeandgraphene} we explore the response of a graphene-coated Drude sphere, conceptually representative of a doped semiconductor, by considering the extinction cross-sectional efficiency. The plasma frequency  considered, $\hbar\omega_{\mathrm{p}} = 0.8\ \text{eV}$, overlaps with the considered graphene Fermi energy range. The intraband approximation in Eq.~\eqref{eq:drudegraphene_intra} plainly provides an excellent account of both the position and width of the dipole resonance in the region $\hbar\omega<2\ef$, i.e.,\ outside the region of interband Landau damping; that is, for sufficiently large spheres at sufficiently high graphene Fermi energies. Notably, the hybridization between the bare resonances of the Drude sphere and a graphene-coated vacuum sphere leads to just a single hybridized mode, rather than the familiar construction of a red- and blueshifted bonding and anti-bonding mode. In the joint Drude-graphene case, due to the absence of spatial separation between the induced charge regions in the two materials -- both residing at the sphere surface -- just a single hybridized plasmon is formed. As evident from Eq.~\eqref{eq:drudegraphene_intra} the hybridized resonance is blueshifted compared to the bare resonances. Significant tunability is achievable by varying either the sphere radius -- or, for dynamical purposes, graphene's doping level. The latter scenario could be achieved e.g., by application of an external gating field, with significant retainment of tunability expected\cite{Gong:2010}, even in the presence of a fixed substrate charge-transfer contribution\cite{Kong:2010}.

\subsection{Relation to surface scattering}
At this point we digress briefly from considerations of graphene-coatings, to consider an endearing corollary of Eq.~\eqref{eq:disprel_stat} in the dipole case related to surface scattering and Kreibig's size-dependent damping model. Specifically, suppose that a metallic particle, well-described by Eq.~\eqref{eq:drudeform}, exhibits a slightly increased damping rate $\tilde{\gamma}_{\text{\textsc{d}}} =\gamma_{\text{\textsc{d}}} +\delta\gamma$ near the surface, e.g., due to roughness. We assume that this region is thin, specifically, it is reasonable to take its width as a single plasma-wavelength $w=\vf/\omega_{\mathrm{p}}$. We include this thin region approximately via a surface conductivity $\sigma = \sigma_{\text{bulk}}w$, where $\sigma_{\text{bulk}}$ denotes the bulk Drude conductivity of loss-rate $\tilde{\gamma}_{\text{\textsc{d}}}$. In this case, working from Eq.~\eqref{eq:disprel_stat}, one finds to first order in the loss-rates and in the ratio $w/R$ that the resonance broadens as $\mathrm{Im}(\omega) \simeq  -\tfrac{1}{2}\big[\gamma + (2\delta\gamma/\omega_{\mathrm{p}})(\vf/R)\big]$, which follows exactly the Kreibig form\cite{Kreibig:1985}, $\gamma\rightarrow\gamma+A\vf/R$, with dimensionless damping parameter $A=2\delta\gamma/\omega_{\mathrm{p}}$. From experimental studies, it is well-known that $A$ is on the order of unity\cite{Kreibig:1985} -- with this in mind, we recognize that surface scattering due to a spatial dependence of $\gamma$ is only a minor contributor to the experimentally measured $A$, since $\delta\gamma/\omega_{\mathrm{p}}\ll 1$ for any reasonable imagined $\delta\gamma$. Indeed, it was established by Apell and Penn, using density functional theory, that the primary contributor to $A$ arises from density inhomogeneity in the surface region\cite{Apell:1983}.

\section{Summary}\label{sec:summary}
In this paper we have examined the electromagnetic response -- and, as a key element, the plasmonic properties -- of a two-component spherical structure, coated at the interface by a conductive film, exemplified here by a graphene coating. Within the naturally suited frame-work of vector waves we derived a corresponding set of generalized Mie--Lorenz scattering coefficients. Restricting our considerations to the quasistatic regime, we derived mathematically uncomplicated dispersion equations for the multipole plasmons. Considering the modest number of established analytical dispersion equations in graphene plasmonics, we believe that the additional member introduced here offers a complementing view, particularly in the emerging subfield of plasmonic interaction in non-planar two-dimensional structures. 
Finally, as useful applications of the theory developed herein, we considered two distinct types of spherical substrates for the coating: specifically, dielectric and Drude substrates. In the former case, this allowed us to explore the localized plasmons arising strictly from the charge-carriers in the graphene coating -- while, in the latter case, we explored the interplay between plasmons supported independently by the bulk and the coating.

\appendix 
\section{Including Hydrodynamic Nonlocality in Graphene's Response in Spherical Geometries}\label{app:hydrodynamic}
We here discuss how to appropriately account for nonlocal response acting in the graphene-coating through a hydrodynamic model. For simplicity -- and because it is justified in the size-regime relevant for nonlocal response in finite structures -- we work in the quasistatic regime. 

Hydrodynamic response is included by a modest generalization of the usual Ohm's law $\mathbf{K}(\rv)=\sigma(\omega)\mathbf{E}_{\scriptscriptstyle\parallel}(\rv)$ by appending to the left-hand side a term $\tfrac{\beta^2}{\omega^2}\nabla_{\scriptscriptstyle\parallel}[\nabla_{\scriptscriptstyle\parallel}\cdot \mathbf{K}(\rv)]$, which in turn, combined with the continuity equation, enforces a relationship between the induced charge density $\rho$ and the potential $\phi$\cite{Wang:2013,Christensen:2014prb}
\begin{equation}\label{eq:hydrodynamiceq}
\rho(\rv)+\frac{\beta^2}{\omega^2}\nabla_{\scriptscriptstyle\parallel}^2\rho(\rv) = \frac{i\sigma(\omega)}{\omega}\nabla_{\scriptscriptstyle\parallel}^2\phi(\rv),
\end{equation}
with plasma velocity $\beta$ proportional to the Fermi velocity $v_{\text{\textsc{f}}}$, interrelated approximately\cite{Christensen:2014prb} by $\beta^2 = {\tfrac{3}{4}}v_{\text{\textsc{f}}}^2$ in graphene.
For the potential, we expand it in the exterior and interior regions, $\mathcal{V}_2$ and $\mathcal{V}_1$, by making use of axial symmetry to freely choose the azimuthal $m=0$ component of a complete set of bounded, non-constant solutions of the Laplace equation\cite{Jackson}
\begin{subequations}\label{eq:potential}
\begin{align} \label{eq:potential_in}
\phi_{\scriptscriptstyle\mathcal{V}_{1}}(\rv) &= \sum_{l=1} c_l^{\mathrm{tra}} r^l P_l(\cos\theta), 
& r < R, \\ \label{eq:potential_out}
\phi_{\scriptscriptstyle\mathcal{V}_{2}}(\rv) &= \sum_{l=1} \big[c_l^{\mathrm{inc}} r^l + c_l^{\mathrm{sca}}r^{-(l+1)}\big] P_l(\cos\theta),
& r > R,
\end{align}
\end{subequations}
with associated incident, scattered, and transmitted multipole coefficients $c_l^{\mathrm{inc}}$, $c_l^{\mathrm{sca}}$, and $c_l^{\mathrm{tra}}$, respectively. Due to spherical symmetry, the coefficients can be matched multipole-by-multipole, i.e., separately for each $l$. The matching is governed by the BCs $\hat{\mathbf{n}}\times(\Ev_{\scriptscriptstyle \mathcal{V}_2}-\Ev_{\scriptscriptstyle \mathcal{V}_1}) = 0$ and $\hat{\mathbf{n}}\times(\mathbf{D}_{\scriptscriptstyle \mathcal{V}_2}-\mathbf{D}_{\scriptscriptstyle \mathcal{V}_1}) = \rho$, which translate into BCs for the potential $\phi$, reading as 
$\partial_\theta\phi_{\scriptscriptstyle\mathcal{V}_2} = \partial_\theta\phi_{\scriptscriptstyle\mathcal{V}_1}$ 
and 
$\varepsilon_1\partial_r\phi_{\scriptscriptstyle\mathcal{V}_1}-\varepsilon_2\partial_r\phi_{\scriptscriptstyle\mathcal{V}_2}  = \rho$ at all surficial points. The induced charge density associated with a potential $\phi_{\scriptscriptstyle \mathcal{V}_1}$ of multipole order $l$ is denoted $\rho_l$ and is obtained by solving Eq.~\eqref{eq:hydrodynamiceq} subject to Eq.~\eqref{eq:potential_in} for fixed $l$ yielding
\begin{subequations}
\begin{equation}\label{eq:surfcharge}
\rho_l = - c_l^{\mathrm{tra}} \frac{i
\sigma_l^{\text{\textsc{h}}}(\omega)}{\omega} R^{l-2} l(l+1)P_l(\cos\theta) ,
\end{equation}
expressed via a hydrodynamically corrected conductivity 
\begin{equation}\label{eq:effectiveconductivity}
\sigma_{l}^{\text{\textsc{h}}}(\omega) \equiv \frac{\sigma(\omega)}{1-\tfrac{\beta^2}{\omega^2}\tfrac{l(l+1)}{R^2}}.
\end{equation}
\end{subequations}
Applying the BCs to Eqs.~\eqref{eq:potential} and~\eqref{eq:surfcharge} then yields a direct relation between the scattered and incident multipole coefficients
\begin{equation}
c_l^{\mathrm{sca}} = -(4\pi)^{-1}\alpha_l^{\text{\textsc{h}}} c_l^{\mathrm{inc}},
\end{equation}
expressed in terms of a hydrodynamic multipole polarizability $\alpha_l^{\text{\textsc{h}}}$. Importantly, as is evident from Eq.~\eqref{eq:surfcharge}, the inclusion of hydrodynamic response acts only to introduce an effective conductivity $\sigma_l^{\text{\textsc{h}}}$. As such, the hydrodynamic multipole polarizability $\alpha_l^{\text{\textsc{h}}}$ differs only from its LRA counterpart $\alpha_l$ of Eq.~\eqref{eq:alpha_l} by the substitution $\sigma\rightarrow\sigma_l^{\text{\textsc{h}}}$. 

Interestingly, in momentum-space the hydrodynamic conductivity of a planar sheet takes the form $\sigma(q,\omega) = \sigma(\omega)\big[1-\tfrac{\beta^2}{\omega^2} q^2\big]^{-1}$. Clearly, a mapping between the planar case and Eq.~\eqref{eq:effectiveconductivity} can be achieved by introducing an effective momentum $q_l^{\text{eff,\textsc{h}}} \equiv \sqrt{l(l+1)}/R$. Notably, this differs from the optically relevant effective momentum $q_l^{\text{eff}}$ at order $\mathcal{O}(l^{-1})$.

Concluding our considerations of hydrodynamics, we comment that the effective nonlocal interaction range $\beta/\omega$ is $\sim\!  1\ \text{nm}$ for graphene (for a resonance e.g., at $\hbar\omega = 0.5\ \text{eV}$). As such, significant hydrodynamic perturbations of the LRA predictions are expected in the proximity of the few-nanometer domain.

\vskip 0.3em
\begin{acknowledgments}
T.C. expresses his gratitude to Weihua~Wang for surfacing the idea of considering a spherical structure, and to Wei~Yan for encouraging and stimulating discussions. The Center for Nanostructured Graphene is sponsored by the Danish National Research Foundation, Project DNRF58.
This work was also supported by the Danish Council for Independent Research, Project 1323-00087.
\end{acknowledgments}


%


\begin{thebibliography}{49}%
\makeatletter
\providecommand \@ifxundefined [1]{%
 \@ifx{#1\undefined}
}%
\providecommand \@ifnum [1]{%
 \ifnum #1\expandafter \@firstoftwo
 \else \expandafter \@secondoftwo
 \fi
}%
\providecommand \@ifx [1]{%
 \ifx #1\expandafter \@firstoftwo
 \else \expandafter \@secondoftwo
 \fi
}%
\providecommand \natexlab [1]{#1}%
\providecommand \enquote  [1]{``#1''}%
\providecommand \bibnamefont  [1]{#1}%
\providecommand \bibfnamefont [1]{#1}%
\providecommand \citenamefont [1]{#1}%
\providecommand \href@noop [0]{\@secondoftwo}%
\providecommand \href [0]{\begingroup \@sanitize@url \@href}%
\providecommand \@href[1]{\@@startlink{#1}\@@href}%
\providecommand \@@href[1]{\endgroup#1\@@endlink}%
\providecommand \@sanitize@url [0]{\catcode `\\12\catcode `\$12\catcode
  `\&12\catcode `\#12\catcode `\^12\catcode `\_12\catcode `\%12\relax}%
\providecommand \@@startlink[1]{}%
\providecommand \@@endlink[0]{}%
\providecommand \url  [0]{\begingroup\@sanitize@url \@url }%
\providecommand \@url [1]{\endgroup\@href {#1}{\urlprefix }}%
\providecommand \urlprefix  [0]{URL }%
\providecommand \Eprint [0]{\href }%
\providecommand \doibase [0]{http://dx.doi.org/}%
\providecommand \selectlanguage [0]{\@gobble}%
\providecommand \bibinfo  [0]{\@secondoftwo}%
\providecommand \bibfield  [0]{\@secondoftwo}%
\providecommand \translation [1]{[#1]}%
\providecommand \BibitemOpen [0]{}%
\providecommand \bibitemStop [0]{}%
\providecommand \bibitemNoStop [0]{.\EOS\space}%
\providecommand \EOS [0]{\spacefactor3000\relax}%
\providecommand \BibitemShut  [1]{\csname bibitem#1\endcsname}%
\let\auto@bib@innerbib\@empty
\bibitem [{\citenamefont {Grigorenko}\ \emph {et~al.}(2012)\citenamefont
  {Grigorenko}, \citenamefont {Polini},\ and\ \citenamefont
  {Novoselov}}]{Grigorenko:2012}%
  \BibitemOpen
  \bibfield  {author} {\bibinfo {author} {\bibfnamefont {{\relax
  A.N}.}~\bibnamefont {Grigorenko}}, \bibinfo {author} {\bibfnamefont
  {M.}~\bibnamefont {Polini}}, \ and\ \bibinfo {author} {\bibfnamefont {{\relax
  K.S}.}~\bibnamefont {Novoselov}},\ }\href {\doibase 10.1038/nphoton.2012.262}
  {\bibfield  {journal} {\bibinfo  {journal} {Nat. Photonics}\ }\textbf
  {\bibinfo {volume} {6}},\ \bibinfo {pages} {749} (\bibinfo {year}
  {2012})}\BibitemShut {NoStop}%
\bibitem [{\citenamefont {Bludov}\ \emph {et~al.}(2013)\citenamefont {Bludov},
  \citenamefont {Ferreira}, \citenamefont {Peres},\ and\ \citenamefont
  {Vasilevskiy}}]{Bludov:2013}%
  \BibitemOpen
  \bibfield  {author} {\bibinfo {author} {\bibfnamefont {{\relax
  Y.V}.}~\bibnamefont {Bludov}}, \bibinfo {author} {\bibfnamefont
  {A.}~\bibnamefont {Ferreira}}, \bibinfo {author} {\bibfnamefont {{\relax
  N.M.R}.}~\bibnamefont {Peres}}, \ and\ \bibinfo {author} {\bibfnamefont
  {{\relax M.I}.}~\bibnamefont {Vasilevskiy}},\ }\href {\doibase
  10.1142/S0217979213410014} {\bibfield  {journal} {\bibinfo  {journal} {Int.
  J. Mod. Phys. B}\ }\textbf {\bibinfo {volume} {27}},\ \bibinfo {pages}
  {1341001} (\bibinfo {year} {2013})}\BibitemShut {NoStop}%
\bibitem [{\citenamefont {Garc\'{i}a~de Abajo}(2014)}]{Abajo:2014}%
  \BibitemOpen
  \bibfield  {author} {\bibinfo {author} {\bibfnamefont {{\relax
  F.J}.}~\bibnamefont {Garc\'{i}a~de Abajo}},\ }\href {\doibase
  10.1021/ph400147y} {\bibfield  {journal} {\bibinfo  {journal} {ACS
  Photonics}\ }\textbf {\bibinfo {volume} {1}},\ \bibinfo {pages} {135}
  (\bibinfo {year} {2014})}\BibitemShut {NoStop}%
\bibitem [{\citenamefont {Wang}\ \emph {et~al.}(2011)\citenamefont {Wang},
  \citenamefont {Apell},\ and\ \citenamefont {Kinaret}}]{Wang:2011}%
  \BibitemOpen
  \bibfield  {author} {\bibinfo {author} {\bibfnamefont {W.}~\bibnamefont
  {Wang}}, \bibinfo {author} {\bibfnamefont {P.}~\bibnamefont {Apell}}, \ and\
  \bibinfo {author} {\bibfnamefont {J.}~\bibnamefont {Kinaret}},\ }\href
  {\doibase 10.1103/PhysRevB.84.085423} {\bibfield  {journal} {\bibinfo
  {journal} {Phys. Rev. B}\ }\textbf {\bibinfo {volume} {84}},\ \bibinfo
  {pages} {085423} (\bibinfo {year} {2011})}\BibitemShut {NoStop}%
\bibitem [{\citenamefont {Chen}\ \emph
  {et~al.}(2012{\natexlab{a}})\citenamefont {Chen}, \citenamefont {Badioli},
  \citenamefont {Alonso-Gonz{\'a}lez}, \citenamefont {Thongrattanasiri},
  \citenamefont {Huth}, \citenamefont {Osmond}, \citenamefont {Spasenovi{\'c}},
  \citenamefont {Centeno}, \citenamefont {Pesquera}, \citenamefont {Godignon},
  \citenamefont {Elorza}, \citenamefont {Camara}, \citenamefont {Garc\'{i}a~de
  Abajo}, \citenamefont {Hillenbrand},\ and\ \citenamefont
  {Koppens}}]{Chen:2012}%
  \BibitemOpen
  \bibfield  {author} {\bibinfo {author} {\bibfnamefont {J.}~\bibnamefont
  {Chen}}, \bibinfo {author} {\bibfnamefont {M.}~\bibnamefont {Badioli}},
  \bibinfo {author} {\bibfnamefont {P.}~\bibnamefont {Alonso-Gonz{\'a}lez}},
  \bibinfo {author} {\bibfnamefont {S.}~\bibnamefont {Thongrattanasiri}},
  \bibinfo {author} {\bibfnamefont {F.}~\bibnamefont {Huth}}, \bibinfo {author}
  {\bibfnamefont {J.}~\bibnamefont {Osmond}}, \bibinfo {author} {\bibfnamefont
  {M.}~\bibnamefont {Spasenovi{\'c}}}, \bibinfo {author} {\bibfnamefont
  {A.}~\bibnamefont {Centeno}}, \bibinfo {author} {\bibfnamefont
  {A.}~\bibnamefont {Pesquera}}, \bibinfo {author} {\bibfnamefont
  {P.}~\bibnamefont {Godignon}}, \bibinfo {author} {\bibfnamefont {{\relax
  A.Z}.}~\bibnamefont {Elorza}}, \bibinfo {author} {\bibfnamefont
  {N.}~\bibnamefont {Camara}}, \bibinfo {author} {\bibfnamefont {{\relax
  F.J}.}~\bibnamefont {Garc\'{i}a~de Abajo}}, \bibinfo {author} {\bibfnamefont
  {R.}~\bibnamefont {Hillenbrand}}, \ and\ \bibinfo {author} {\bibfnamefont
  {{\relax F.H.L}.}~\bibnamefont {Koppens}},\ }\href {\doibase
  10.1038/nature11254} {\bibfield  {journal} {\bibinfo  {journal} {Nature}\
  }\textbf {\bibinfo {volume} {487}},\ \bibinfo {pages} {77} (\bibinfo {year}
  {2012}{\natexlab{a}})}\BibitemShut {NoStop}%
\bibitem [{\citenamefont {Fei}\ \emph {et~al.}(2012)\citenamefont {Fei},
  \citenamefont {Rodin}, \citenamefont {Andreev}, \citenamefont {Bao},
  \citenamefont {McLeod}, \citenamefont {Wagner}, \citenamefont {Zhang},
  \citenamefont {Zhao}, \citenamefont {Thiemens}, \citenamefont {Dominguez},
  \citenamefont {Fogler}, \citenamefont {Castro~Neto}, \citenamefont {Lau},
  \citenamefont {Keilmann},\ and\ \citenamefont {Basov}}]{Fei:2012}%
  \BibitemOpen
  \bibfield  {author} {\bibinfo {author} {\bibfnamefont {Z.}~\bibnamefont
  {Fei}}, \bibinfo {author} {\bibfnamefont {A.}~\bibnamefont {Rodin}}, \bibinfo
  {author} {\bibfnamefont {G.}~\bibnamefont {Andreev}}, \bibinfo {author}
  {\bibfnamefont {W.}~\bibnamefont {Bao}}, \bibinfo {author} {\bibfnamefont
  {A.}~\bibnamefont {McLeod}}, \bibinfo {author} {\bibfnamefont
  {M.}~\bibnamefont {Wagner}}, \bibinfo {author} {\bibfnamefont
  {L.}~\bibnamefont {Zhang}}, \bibinfo {author} {\bibfnamefont
  {Z.}~\bibnamefont {Zhao}}, \bibinfo {author} {\bibfnamefont {M.}~\bibnamefont
  {Thiemens}}, \bibinfo {author} {\bibfnamefont {G.}~\bibnamefont {Dominguez}},
  \bibinfo {author} {\bibfnamefont {{\relax M.M}.}~\bibnamefont {Fogler}},
  \bibinfo {author} {\bibfnamefont {{\relax A.H}.}~\bibnamefont {Castro~Neto}},
  \bibinfo {author} {\bibfnamefont {{\relax C.N}.}~\bibnamefont {Lau}},
  \bibinfo {author} {\bibfnamefont {F.}~\bibnamefont {Keilmann}}, \ and\
  \bibinfo {author} {\bibfnamefont {{\relax D.N}.}~\bibnamefont {Basov}},\
  }\href {\doibase 10.1038/nature11253} {\bibfield  {journal} {\bibinfo
  {journal} {Nature}\ }\textbf {\bibinfo {volume} {487}},\ \bibinfo {pages}
  {82} (\bibinfo {year} {2012})}\BibitemShut {NoStop}%
\bibitem [{\citenamefont {Thongrattanasiri}\ and\ \citenamefont {Garc\'{i}a~de
  Abajo}(2013)}]{Thongrattanasiri:2013a}%
  \BibitemOpen
  \bibfield  {author} {\bibinfo {author} {\bibfnamefont {S.}~\bibnamefont
  {Thongrattanasiri}}\ and\ \bibinfo {author} {\bibfnamefont {{\relax
  F.J}.}~\bibnamefont {Garc\'{i}a~de Abajo}},\ }\href {\doibase
  10.1103/PhysRevLett.110.187401} {\bibfield  {journal} {\bibinfo  {journal}
  {Phys. Rev. Lett.}\ }\textbf {\bibinfo {volume} {110}},\ \bibinfo {pages}
  {187401} (\bibinfo {year} {2013})}\BibitemShut {NoStop}%
\bibitem [{\citenamefont {Thongrattanasiri}\ \emph
  {et~al.}(2012{\natexlab{a}})\citenamefont {Thongrattanasiri}, \citenamefont
  {Koppens},\ and\ \citenamefont {Garc\'{i}a~de
  Abajo}}]{Thongrattanasiri:2012a}%
  \BibitemOpen
  \bibfield  {author} {\bibinfo {author} {\bibfnamefont {S.}~\bibnamefont
  {Thongrattanasiri}}, \bibinfo {author} {\bibfnamefont {{\relax
  F.H.L}.}~\bibnamefont {Koppens}}, \ and\ \bibinfo {author} {\bibfnamefont
  {{\relax F.J}.}~\bibnamefont {Garc\'{i}a~de Abajo}},\ }\href {\doibase
  10.1103/PhysRevLett.108.047401} {\bibfield  {journal} {\bibinfo  {journal}
  {Phys. Rev. Lett.}\ }\textbf {\bibinfo {volume} {108}},\ \bibinfo {pages}
  {047401} (\bibinfo {year} {2012}{\natexlab{a}})}\BibitemShut {NoStop}%
\bibitem [{\citenamefont {Fang}\ \emph {et~al.}(2012)\citenamefont {Fang},
  \citenamefont {Thongrattanasiri}, \citenamefont {Schlather}, \citenamefont
  {Liu}, \citenamefont {Ma}, \citenamefont {Wang}, \citenamefont {Ajayan},
  \citenamefont {Nordlander}, \citenamefont {Halas},\ and\ \citenamefont
  {Garc\'{i}a~de Abajo}}]{Fang:2012}%
  \BibitemOpen
  \bibfield  {author} {\bibinfo {author} {\bibfnamefont {Z.}~\bibnamefont
  {Fang}}, \bibinfo {author} {\bibfnamefont {S.}~\bibnamefont
  {Thongrattanasiri}}, \bibinfo {author} {\bibfnamefont {A.}~\bibnamefont
  {Schlather}}, \bibinfo {author} {\bibfnamefont {Z.}~\bibnamefont {Liu}},
  \bibinfo {author} {\bibfnamefont {L.}~\bibnamefont {Ma}}, \bibinfo {author}
  {\bibfnamefont {Y.}~\bibnamefont {Wang}}, \bibinfo {author} {\bibfnamefont
  {{\relax P.M}.}~\bibnamefont {Ajayan}}, \bibinfo {author} {\bibfnamefont
  {P.}~\bibnamefont {Nordlander}}, \bibinfo {author} {\bibfnamefont {{\relax
  N.J}.}~\bibnamefont {Halas}}, \ and\ \bibinfo {author} {\bibfnamefont
  {{\relax F.J}.}~\bibnamefont {Garc\'{i}a~de Abajo}},\ }\href {\doibase
  10.1021/nn3055835} {\bibfield  {journal} {\bibinfo  {journal} {ACS Nano}\
  }\textbf {\bibinfo {volume} {7}},\ \bibinfo {pages} {2388} (\bibinfo {year}
  {2012})}\BibitemShut {NoStop}%
\bibitem [{\citenamefont {Zhu}\ \emph {et~al.}(2014{\natexlab{a}})\citenamefont
  {Zhu}, \citenamefont {Wang}, \citenamefont {Yan}, \citenamefont {Larsen},
  \citenamefont {B\o{}ggild}, \citenamefont {Pedersen}, \citenamefont {Xiao},
  \citenamefont {Zi},\ and\ \citenamefont {Mortensen}}]{XZhu:2014}%
  \BibitemOpen
  \bibfield  {author} {\bibinfo {author} {\bibfnamefont {X.}~\bibnamefont
  {Zhu}}, \bibinfo {author} {\bibfnamefont {W.}~\bibnamefont {Wang}}, \bibinfo
  {author} {\bibfnamefont {W.}~\bibnamefont {Yan}}, \bibinfo {author}
  {\bibfnamefont {{\relax M.B}.}~\bibnamefont {Larsen}}, \bibinfo {author}
  {\bibfnamefont {P.}~\bibnamefont {B\o{}ggild}}, \bibinfo {author}
  {\bibfnamefont {{\relax T.G}.}~\bibnamefont {Pedersen}}, \bibinfo {author}
  {\bibfnamefont {S.}~\bibnamefont {Xiao}}, \bibinfo {author} {\bibfnamefont
  {J.}~\bibnamefont {Zi}}, \ and\ \bibinfo {author} {\bibfnamefont {{\relax
  N.A}.}~\bibnamefont {Mortensen}},\ }\href {\doibase 10.1021/nl500948p}
  {\bibfield  {journal} {\bibinfo  {journal} {Nano Lett.}\ }\textbf {\bibinfo
  {volume} {14}},\ \bibinfo {pages} {2907} (\bibinfo {year}
  {2014}{\natexlab{a}})}\BibitemShut {NoStop}%
\bibitem [{\citenamefont {Lu}\ \emph {et~al.}(2013)\citenamefont {Lu},
  \citenamefont {Zhu}, \citenamefont {Xu}, \citenamefont {Ni}, \citenamefont
  {Dong},\ and\ \citenamefont {Cui}}]{BingLu:2013}%
  \BibitemOpen
  \bibfield  {author} {\bibinfo {author} {\bibfnamefont {W.~B.}\ \bibnamefont
  {Lu}}, \bibinfo {author} {\bibfnamefont {W.}~\bibnamefont {Zhu}}, \bibinfo
  {author} {\bibfnamefont {H.~J.}\ \bibnamefont {Xu}}, \bibinfo {author}
  {\bibfnamefont {Z.~H.}\ \bibnamefont {Ni}}, \bibinfo {author} {\bibfnamefont
  {Z.~G.}\ \bibnamefont {Dong}}, \ and\ \bibinfo {author} {\bibfnamefont
  {T.~J.}\ \bibnamefont {Cui}},\ }\href {\doibase 10.1364/OE.21.010475}
  {\bibfield  {journal} {\bibinfo  {journal} {Opt. Express}\ }\textbf {\bibinfo
  {volume} {21}},\ \bibinfo {pages} {10475} (\bibinfo {year}
  {2013})}\BibitemShut {NoStop}%
\bibitem [{\citenamefont {Chen}\ \emph
  {et~al.}(2012{\natexlab{b}})\citenamefont {Chen}, \citenamefont {Soric},\
  and\ \citenamefont {Al\'{u}}}]{ChenAlu:2012}%
  \BibitemOpen
  \bibfield  {author} {\bibinfo {author} {\bibfnamefont {{\relax
  P.-Y}.}~\bibnamefont {Chen}}, \bibinfo {author} {\bibfnamefont
  {J.}~\bibnamefont {Soric}}, \ and\ \bibinfo {author} {\bibfnamefont
  {A.}~\bibnamefont {Al\'{u}}},\ }\href {\doibase 10.1002/adma.201202624}
  {\bibfield  {journal} {\bibinfo  {journal} {Adv. Mater.}\ }\textbf {\bibinfo
  {volume} {24}},\ \bibinfo {pages} {OP281} (\bibinfo {year}
  {2012}{\natexlab{b}})}\BibitemShut {NoStop}%
\bibitem [{\citenamefont {Chen}\ \emph {et~al.}(2013)\citenamefont {Chen},
  \citenamefont {Soric}, \citenamefont {Padooru}, \citenamefont {Bernety},
  \citenamefont {Yakovlev},\ and\ \citenamefont {Al\'{u}}}]{ChenAlu:2013}%
  \BibitemOpen
  \bibfield  {author} {\bibinfo {author} {\bibfnamefont {P.}~\bibnamefont
  {Chen}}, \bibinfo {author} {\bibfnamefont {J.}~\bibnamefont {Soric}},
  \bibinfo {author} {\bibfnamefont {{\relax Y.R}.}~\bibnamefont {Padooru}},
  \bibinfo {author} {\bibfnamefont {{\relax H.M}.}~\bibnamefont {Bernety}},
  \bibinfo {author} {\bibfnamefont {{\relax A.B}.}~\bibnamefont {Yakovlev}}, \
  and\ \bibinfo {author} {\bibfnamefont {A.}~\bibnamefont {Al\'{u}}},\ }\href
  {\doibase 10.1088/1367-2630/15/12/123029} {\bibfield  {journal} {\bibinfo
  {journal} {New J. Phys.}\ }\textbf {\bibinfo {volume} {15}},\ \bibinfo
  {pages} {123029} (\bibinfo {year} {2013})}\BibitemShut {NoStop}%
\bibitem [{\citenamefont {Zhu}\ \emph {et~al.}(2014{\natexlab{b}})\citenamefont
  {Zhu}, \citenamefont {Ren}, \citenamefont {Gao}, \citenamefont {Yang},
  \citenamefont {Lian},\ and\ \citenamefont {Jian}}]{Zhu:2014}%
  \BibitemOpen
  \bibfield  {author} {\bibinfo {author} {\bibfnamefont {B.}~\bibnamefont
  {Zhu}}, \bibinfo {author} {\bibfnamefont {G.}~\bibnamefont {Ren}}, \bibinfo
  {author} {\bibfnamefont {Y.}~\bibnamefont {Gao}}, \bibinfo {author}
  {\bibfnamefont {Y.}~\bibnamefont {Yang}}, \bibinfo {author} {\bibfnamefont
  {Y.}~\bibnamefont {Lian}}, \ and\ \bibinfo {author} {\bibfnamefont
  {S.}~\bibnamefont {Jian}},\ }\href {\doibase 10.1364/OE.22.024096} {\bibfield
   {journal} {\bibinfo  {journal} {Opt. Express}\ }\textbf {\bibinfo {volume}
  {22}},\ \bibinfo {pages} {24096} (\bibinfo {year}
  {2014}{\natexlab{b}})}\BibitemShut {NoStop}%
\bibitem [{\citenamefont {Gao}\ \emph {et~al.}(2014{\natexlab{a}})\citenamefont
  {Gao}, \citenamefont {Ren}, \citenamefont {Zhu}, \citenamefont {Liu},
  \citenamefont {Lian},\ and\ \citenamefont {Jian}}]{Gao:2014a}%
  \BibitemOpen
  \bibfield  {author} {\bibinfo {author} {\bibfnamefont {Y.}~\bibnamefont
  {Gao}}, \bibinfo {author} {\bibfnamefont {G.}~\bibnamefont {Ren}}, \bibinfo
  {author} {\bibfnamefont {B.}~\bibnamefont {Zhu}}, \bibinfo {author}
  {\bibfnamefont {H.}~\bibnamefont {Liu}}, \bibinfo {author} {\bibfnamefont
  {Y.}~\bibnamefont {Lian}}, \ and\ \bibinfo {author} {\bibfnamefont
  {S.}~\bibnamefont {Jian}},\ }\href {\doibase 10.1364/OE.22.024322} {\bibfield
   {journal} {\bibinfo  {journal} {Opt. Express}\ }\textbf {\bibinfo {volume}
  {22}},\ \bibinfo {pages} {24322} (\bibinfo {year}
  {2014}{\natexlab{a}})}\BibitemShut {NoStop}%
\bibitem [{\citenamefont {Gao}\ \emph {et~al.}(2014{\natexlab{b}})\citenamefont
  {Gao}, \citenamefont {Ren}, \citenamefont {Zhu}, \citenamefont {Wang},\ and\
  \citenamefont {Jian}}]{Gao:2014b}%
  \BibitemOpen
  \bibfield  {author} {\bibinfo {author} {\bibfnamefont {Y.}~\bibnamefont
  {Gao}}, \bibinfo {author} {\bibfnamefont {G.}~\bibnamefont {Ren}}, \bibinfo
  {author} {\bibfnamefont {B.}~\bibnamefont {Zhu}}, \bibinfo {author}
  {\bibfnamefont {J.}~\bibnamefont {Wang}}, \ and\ \bibinfo {author}
  {\bibfnamefont {S.}~\bibnamefont {Jian}},\ }\href {\doibase
  10.1364/OL.39.005909} {\bibfield  {journal} {\bibinfo  {journal} {Opt.
  Lett.}\ }\textbf {\bibinfo {volume} {39}},\ \bibinfo {pages} {5909} (\bibinfo
  {year} {2014}{\natexlab{b}})}\BibitemShut {NoStop}%
\bibitem [{\citenamefont {Huang}\ \emph {et~al.}(2014)\citenamefont {Huang},
  \citenamefont {Wang}, \citenamefont {Sun}, \citenamefont {He}, \citenamefont
  {Liu}, \citenamefont {Li},\ and\ \citenamefont {Zhai}}]{Huang:2014}%
  \BibitemOpen
  \bibfield  {author} {\bibinfo {author} {\bibfnamefont {Z.-R.}\ \bibnamefont
  {Huang}}, \bibinfo {author} {\bibfnamefont {L.-L.}\ \bibnamefont {Wang}},
  \bibinfo {author} {\bibfnamefont {B.}~\bibnamefont {Sun}}, \bibinfo {author}
  {\bibfnamefont {M.-D.}\ \bibnamefont {He}}, \bibinfo {author} {\bibfnamefont
  {J.-Q.}\ \bibnamefont {Liu}}, \bibinfo {author} {\bibfnamefont {H.-J.}\
  \bibnamefont {Li}}, \ and\ \bibinfo {author} {\bibfnamefont {X.}~\bibnamefont
  {Zhai}},\ }\href {http://stacks.iop.org/2040-8986/16/i=10/a=105004}
  {\bibfield  {journal} {\bibinfo  {journal} {J. Opt.}\ }\textbf {\bibinfo
  {volume} {16}},\ \bibinfo {pages} {105004} (\bibinfo {year}
  {2014})}\BibitemShut {NoStop}%
\bibitem [{\citenamefont {Yang}\ \emph {et~al.}(2014)\citenamefont {Yang},
  \citenamefont {Hou}, \citenamefont {Zhou}, \citenamefont {He}, \citenamefont
  {Cao},\ and\ \citenamefont {Kuang}}]{Yang:2014}%
  \BibitemOpen
  \bibfield  {author} {\bibinfo {author} {\bibfnamefont {H.}~\bibnamefont
  {Yang}}, \bibinfo {author} {\bibfnamefont {Z.}~\bibnamefont {Hou}}, \bibinfo
  {author} {\bibfnamefont {N.}~\bibnamefont {Zhou}}, \bibinfo {author}
  {\bibfnamefont {B.}~\bibnamefont {He}}, \bibinfo {author} {\bibfnamefont
  {J.}~\bibnamefont {Cao}}, \ and\ \bibinfo {author} {\bibfnamefont
  {Y.}~\bibnamefont {Kuang}},\ }\href {\doibase 10.1016/j.ceramint.2014.05.109}
  {\bibfield  {journal} {\bibinfo  {journal} {Ceramics International}\ }\textbf
  {\bibinfo {volume} {40}},\ \bibinfo {pages} {13903} (\bibinfo {year}
  {2014})}\BibitemShut {NoStop}%
\bibitem [{\citenamefont {Lee}\ \emph {et~al.}(2013)\citenamefont {Lee},
  \citenamefont {Kim}, \citenamefont {Yoon},\ and\ \citenamefont
  {Jang}}]{Lee:2013}%
  \BibitemOpen
  \bibfield  {author} {\bibinfo {author} {\bibfnamefont {{\relax
  J.-S}.}~\bibnamefont {Lee}}, \bibinfo {author} {\bibfnamefont {{\relax
  S.-I}.}~\bibnamefont {Kim}}, \bibinfo {author} {\bibfnamefont {{\relax
  J.-C}.}~\bibnamefont {Yoon}}, \ and\ \bibinfo {author} {\bibfnamefont
  {{\relax J.-H}.}~\bibnamefont {Jang}},\ }\href {\doibase 10.1021/nn401850z}
  {\bibfield  {journal} {\bibinfo  {journal} {ACS Nano}\ }\textbf {\bibinfo
  {volume} {7}},\ \bibinfo {pages} {6047} (\bibinfo {year} {2013})}\BibitemShut
  {NoStop}%
\bibitem [{\citenamefont {Bakowies}\ \emph {et~al.}(1995)\citenamefont
  {Bakowies}, \citenamefont {Buehl},\ and\ \citenamefont
  {Thiel}}]{Bakowies:1995}%
  \BibitemOpen
  \bibfield  {author} {\bibinfo {author} {\bibfnamefont {D.}~\bibnamefont
  {Bakowies}}, \bibinfo {author} {\bibfnamefont {M.}~\bibnamefont {Buehl}}, \
  and\ \bibinfo {author} {\bibfnamefont {W.}~\bibnamefont {Thiel}},\ }\href
  {\doibase 10.1021/ja00145a025} {\bibfield  {journal} {\bibinfo  {journal} {J.
  Am. Chem. Soc.}\ }\textbf {\bibinfo {volume} {117}},\ \bibinfo {pages}
  {10113} (\bibinfo {year} {1995})}\BibitemShut {NoStop}%
\bibitem [{\citenamefont {Calaminici}\ \emph {et~al.}(2009)\citenamefont
  {Calaminici}, \citenamefont {Geudtner},\ and\ \citenamefont
  {K\"{o}ster}}]{Calaminici:2009}%
  \BibitemOpen
  \bibfield  {author} {\bibinfo {author} {\bibfnamefont {P.}~\bibnamefont
  {Calaminici}}, \bibinfo {author} {\bibfnamefont {G.}~\bibnamefont
  {Geudtner}}, \ and\ \bibinfo {author} {\bibfnamefont {A.~M.}\ \bibnamefont
  {K\"{o}ster}},\ }\href {\doibase 10.1021/ct800347u} {\bibfield  {journal}
  {\bibinfo  {journal} {J. Chem. Theory Comput.}\ }\textbf {\bibinfo {volume}
  {5}},\ \bibinfo {pages} {29} (\bibinfo {year} {2009})}\BibitemShut {NoStop}%
\bibitem [{\citenamefont {Thongrattanasiri}\ \emph
  {et~al.}(2012{\natexlab{b}})\citenamefont {Thongrattanasiri}, \citenamefont
  {Manjavacas},\ and\ \citenamefont {Garc\'{i}a~de
  Abajo}}]{Thongrattanasiri:2012}%
  \BibitemOpen
  \bibfield  {author} {\bibinfo {author} {\bibfnamefont {S.}~\bibnamefont
  {Thongrattanasiri}}, \bibinfo {author} {\bibfnamefont {A.}~\bibnamefont
  {Manjavacas}}, \ and\ \bibinfo {author} {\bibfnamefont {{\relax
  F.J}.}~\bibnamefont {Garc\'{i}a~de Abajo}},\ }\href {\doibase
  10.1021/nn204780e} {\bibfield  {journal} {\bibinfo  {journal} {ACS Nano}\
  }\textbf {\bibinfo {volume} {6}},\ \bibinfo {pages} {1766} (\bibinfo {year}
  {2012}{\natexlab{b}})}\BibitemShut {NoStop}%
\bibitem [{\citenamefont {Stratton}(1941)}]{Stratton}%
  \BibitemOpen
  \bibfield  {author} {\bibinfo {author} {\bibfnamefont {J.~A.}\ \bibnamefont
  {Stratton}},\ }\href@noop {} {\emph {\bibinfo {title} {{E}lectromagnetic
  {T}heory}}}\ (\bibinfo  {publisher} {McGraw-Hill Book Company},\ \bibinfo
  {address} {New York},\ \bibinfo {year} {1941})\BibitemShut {NoStop}%
\bibitem [{\citenamefont {Ruppin}(1973)}]{Ruppin:1973a}%
  \BibitemOpen
  \bibfield  {author} {\bibinfo {author} {\bibfnamefont {R.}~\bibnamefont
  {Ruppin}},\ }\href {\doibase 10.1103/PhysRevLett.31.1434} {\bibfield
  {journal} {\bibinfo  {journal} {Phys. Rev. Lett.}\ }\textbf {\bibinfo
  {volume} {31}},\ \bibinfo {pages} {1434} (\bibinfo {year}
  {1973})}\BibitemShut {NoStop}%
\bibitem [{\citenamefont {David}\ and\ \citenamefont {Garc{\' i}a~de
  Abajo}(2012)}]{David:2012}%
  \BibitemOpen
  \bibfield  {author} {\bibinfo {author} {\bibfnamefont {C.}~\bibnamefont
  {David}}\ and\ \bibinfo {author} {\bibfnamefont {F.~J.}\ \bibnamefont
  {Garc{\' i}a~de Abajo}},\ }\href {\doibase 10.1021/jp204261u} {\bibfield
  {journal} {\bibinfo  {journal} {J. Phys. Chem. C}\ }\textbf {\bibinfo
  {volume} {115}},\ \bibinfo {pages} {19470} (\bibinfo {year}
  {2012})}\BibitemShut {NoStop}%
\bibitem [{\citenamefont {Christensen}\ \emph
  {et~al.}(2014{\natexlab{a}})\citenamefont {Christensen}, \citenamefont {Yan},
  \citenamefont {Raza}, \citenamefont {Jauho}, \citenamefont {Mortensen},\ and\
  \citenamefont {Wubs}}]{Christensen:2014}%
  \BibitemOpen
  \bibfield  {author} {\bibinfo {author} {\bibfnamefont {T.}~\bibnamefont
  {Christensen}}, \bibinfo {author} {\bibfnamefont {W.}~\bibnamefont {Yan}},
  \bibinfo {author} {\bibfnamefont {S.}~\bibnamefont {Raza}}, \bibinfo {author}
  {\bibfnamefont {{\relax A.-P}.}~\bibnamefont {Jauho}}, \bibinfo {author}
  {\bibfnamefont {{\relax N.A}.}~\bibnamefont {Mortensen}}, \ and\ \bibinfo
  {author} {\bibfnamefont {M.}~\bibnamefont {Wubs}},\ }\href {\doibase
  10.1021/nn406153k} {\bibfield  {journal} {\bibinfo  {journal} {ACS Nano}\
  }\textbf {\bibinfo {volume} {8}},\ \bibinfo {pages} {1745} (\bibinfo {year}
  {2014}{\natexlab{a}})}\BibitemShut {NoStop}%
\bibitem [{\citenamefont {Bohren}\ and\ \citenamefont
  {Huffman}(1983)}]{BohrenHuffman:1983}%
  \BibitemOpen
  \bibfield  {author} {\bibinfo {author} {\bibfnamefont {C.~F.}\ \bibnamefont
  {Bohren}}\ and\ \bibinfo {author} {\bibfnamefont {D.~R.}\ \bibnamefont
  {Huffman}},\ }\href {\doibase 10.1002/9783527618156} {\emph {\bibinfo {title}
  {{A}bsorption and {S}cattering of {L}ight by {S}mall {P}articles}}}\
  (\bibinfo  {publisher} {John Wiley \& Sons},\ \bibinfo {address} {New York},\
  \bibinfo {year} {1983})\BibitemShut {NoStop}%
\bibitem [{\citenamefont {Koppens}\ \emph {et~al.}(2011)\citenamefont
  {Koppens}, \citenamefont {Chang},\ and\ \citenamefont {Garc\'\i{}a~de
  Abajo}}]{Koppens:2011}%
  \BibitemOpen
  \bibfield  {author} {\bibinfo {author} {\bibfnamefont {{\relax
  F.H.L}.}~\bibnamefont {Koppens}}, \bibinfo {author} {\bibfnamefont {{\relax
  D.E}.}~\bibnamefont {Chang}}, \ and\ \bibinfo {author} {\bibfnamefont
  {{\relax F.J}.}~\bibnamefont {Garc\'\i{}a~de Abajo}},\ }\href {\doibase
  10.1021/nl201771h} {\bibfield  {journal} {\bibinfo  {journal} {Nano Lett.}\
  }\textbf {\bibinfo {volume} {11}},\ \bibinfo {pages} {3370} (\bibinfo {year}
  {2011})}\BibitemShut {NoStop}%
\bibitem [{\citenamefont {Fuchs}\ and\ \citenamefont
  {Claro}(1987)}]{Fuchs:1987}%
  \BibitemOpen
  \bibfield  {author} {\bibinfo {author} {\bibfnamefont {R.}~\bibnamefont
  {Fuchs}}\ and\ \bibinfo {author} {\bibfnamefont {F.}~\bibnamefont {Claro}},\
  }\href {\doibase 10.1103/PhysRevB.35.3722} {\bibfield  {journal} {\bibinfo
  {journal} {Phys. Rev. B}\ }\textbf {\bibinfo {volume} {35}},\ \bibinfo
  {pages} {3722} (\bibinfo {year} {1987})}\BibitemShut {NoStop}%
\bibitem [{\citenamefont {Myroshnychenko}\ \emph {et~al.}(2008)\citenamefont
  {Myroshnychenko}, \citenamefont {Rodr\'{\i}guez-Fern\'{a}ndez}, \citenamefont
  {Pastoriza-Santos}, \citenamefont {Funston}, \citenamefont {Novo},
  \citenamefont {Mulvaney}, \citenamefont {Liz-Marz\'{a}n},\ and\ \citenamefont
  {Garc\'{\i}a~de Abajo}}]{Myroshnychenko:2008}%
  \BibitemOpen
  \bibfield  {author} {\bibinfo {author} {\bibfnamefont {V.}~\bibnamefont
  {Myroshnychenko}}, \bibinfo {author} {\bibfnamefont {J.}~\bibnamefont
  {Rodr\'{\i}guez-Fern\'{a}ndez}}, \bibinfo {author} {\bibfnamefont
  {I.}~\bibnamefont {Pastoriza-Santos}}, \bibinfo {author} {\bibfnamefont
  {{\relax A.M}.}~\bibnamefont {Funston}}, \bibinfo {author} {\bibfnamefont
  {C.}~\bibnamefont {Novo}}, \bibinfo {author} {\bibfnamefont {P.}~\bibnamefont
  {Mulvaney}}, \bibinfo {author} {\bibfnamefont {{\relax L.M}.}~\bibnamefont
  {Liz-Marz\'{a}n}}, \ and\ \bibinfo {author} {\bibfnamefont {{\relax
  F.J}.}~\bibnamefont {Garc\'{\i}a~de Abajo}},\ }\href {\doibase
  10.1039/b711486a} {\bibfield  {journal} {\bibinfo  {journal} {Chem. Soc.
  Rev.}\ }\textbf {\bibinfo {volume} {37}},\ \bibinfo {pages} {1792} (\bibinfo
  {year} {2008})}\BibitemShut {NoStop}%
\bibitem [{\citenamefont {Low}\ and\ \citenamefont {Avouris}(2014)}]{Low:2014}%
  \BibitemOpen
  \bibfield  {author} {\bibinfo {author} {\bibfnamefont {T.}~\bibnamefont
  {Low}}\ and\ \bibinfo {author} {\bibfnamefont {P.}~\bibnamefont {Avouris}},\
  }\href {\doibase 10.1021/nn406627u} {\bibfield  {journal} {\bibinfo
  {journal} {ACS Nano}\ }\textbf {\bibinfo {volume} {8}},\ \bibinfo {pages}
  {1086} (\bibinfo {year} {2014})}\BibitemShut {NoStop}%
\bibitem [{\citenamefont {Wunsch}\ \emph {et~al.}(2006)\citenamefont {Wunsch},
  \citenamefont {Stauber}, \citenamefont {Sols},\ and\ \citenamefont
  {Guinea}}]{Wunsch:2006}%
  \BibitemOpen
  \bibfield  {author} {\bibinfo {author} {\bibfnamefont {B.}~\bibnamefont
  {Wunsch}}, \bibinfo {author} {\bibfnamefont {T.}~\bibnamefont {Stauber}},
  \bibinfo {author} {\bibfnamefont {F.}~\bibnamefont {Sols}}, \ and\ \bibinfo
  {author} {\bibfnamefont {F.}~\bibnamefont {Guinea}},\ }\href {\doibase
  10.1088/1367-2630/8/12/318} {\bibfield  {journal} {\bibinfo  {journal} {New
  J. Phys.}\ }\textbf {\bibinfo {volume} {8}},\ \bibinfo {pages} {318}
  (\bibinfo {year} {2006})}\BibitemShut {NoStop}%
\bibitem [{\citenamefont {Hwang}\ and\ \citenamefont
  {Das~Sarma}(2007)}]{Hwang:2007}%
  \BibitemOpen
  \bibfield  {author} {\bibinfo {author} {\bibfnamefont {{\relax
  E.H}.}~\bibnamefont {Hwang}}\ and\ \bibinfo {author} {\bibfnamefont
  {S.}~\bibnamefont {Das~Sarma}},\ }\href {\doibase 10.1103/PhysRevB.75.205418}
  {\bibfield  {journal} {\bibinfo  {journal} {Phys. Rev. B}\ }\textbf {\bibinfo
  {volume} {75}},\ \bibinfo {pages} {205418} (\bibinfo {year}
  {2007})}\BibitemShut {NoStop}%
\bibitem [{\citenamefont {Falkovsky}\ and\ \citenamefont
  {Varlamov}(2007)}]{Falkovsky:2007}%
  \BibitemOpen
  \bibfield  {author} {\bibinfo {author} {\bibfnamefont {{\relax
  L.A}.}~\bibnamefont {Falkovsky}}\ and\ \bibinfo {author} {\bibfnamefont
  {{\relax A.A}.}~\bibnamefont {Varlamov}},\ }\href {\doibase
  10.1140/epjb/e2007-00142-3} {\bibfield  {journal} {\bibinfo  {journal} {Eur.
  Phys. J. B}\ }\textbf {\bibinfo {volume} {56}},\ \bibinfo {pages} {281}
  (\bibinfo {year} {2007})}\BibitemShut {NoStop}%
\bibitem [{\citenamefont {Gusynin}\ \emph {et~al.}(2007)\citenamefont
  {Gusynin}, \citenamefont {Sharapov},\ and\ \citenamefont
  {Carbotte}}]{Gusynin:2007}%
  \BibitemOpen
  \bibfield  {author} {\bibinfo {author} {\bibfnamefont {{\relax
  V.P}.}~\bibnamefont {Gusynin}}, \bibinfo {author} {\bibfnamefont {{\relax
  S.G}.}~\bibnamefont {Sharapov}}, \ and\ \bibinfo {author} {\bibfnamefont
  {{\relax J.P}.}~\bibnamefont {Carbotte}},\ }\href {\doibase
  10.1088/0953-8984/19/2/026222} {\bibfield  {journal} {\bibinfo  {journal} {J.
  Phys.: Condens. Matter}\ }\textbf {\bibinfo {volume} {19}},\ \bibinfo {pages}
  {026222} (\bibinfo {year} {2007})}\BibitemShut {NoStop}%
\bibitem [{\citenamefont {Hanson}(2008)}]{Hanson:2008}%
  \BibitemOpen
  \bibfield  {author} {\bibinfo {author} {\bibfnamefont {{\relax
  G.W}.}~\bibnamefont {Hanson}},\ }\href {\doibase 10.1063/1.3005881}
  {\bibfield  {journal} {\bibinfo  {journal} {J. Appl. Phys.}\ }\textbf
  {\bibinfo {volume} {104}},\ \bibinfo {pages} {084314} (\bibinfo {year}
  {2008})}\BibitemShut {NoStop}%
\bibitem [{\citenamefont {Wang}\ and\ \citenamefont
  {Kinaret}(2013)}]{Wang:2013}%
  \BibitemOpen
  \bibfield  {author} {\bibinfo {author} {\bibfnamefont {W.}~\bibnamefont
  {Wang}}\ and\ \bibinfo {author} {\bibfnamefont {{\relax J.M}.}~\bibnamefont
  {Kinaret}},\ }\href {\doibase 10.1103/PhysRevB.87.195424} {\bibfield
  {journal} {\bibinfo  {journal} {Phys. Rev. B}\ }\textbf {\bibinfo {volume}
  {87}},\ \bibinfo {pages} {195424} (\bibinfo {year} {2013})}\BibitemShut
  {NoStop}%
\bibitem [{\citenamefont {Mortensen}\ \emph {et~al.}(2014)\citenamefont
  {Mortensen}, \citenamefont {Raza}, \citenamefont {Wubs},\ and\ \citenamefont
  {Bozhevolnyi}}]{Mortensen:2014}%
  \BibitemOpen
  \bibfield  {author} {\bibinfo {author} {\bibfnamefont {{\relax
  N.A}.}~\bibnamefont {Mortensen}}, \bibinfo {author} {\bibfnamefont
  {S.}~\bibnamefont {Raza}}, \bibinfo {author} {\bibfnamefont {M.}~\bibnamefont
  {Wubs}}, \ and\ \bibinfo {author} {\bibfnamefont {S.}~\bibnamefont
  {Bozhevolnyi}},\ }\href {\doibase 10.1038/ncomms4809} {\bibfield  {journal}
  {\bibinfo  {journal} {Nat. Commun.}\ }\textbf {\bibinfo {volume} {5}},\
  \bibinfo {pages} {3809} (\bibinfo {year} {2014})}\BibitemShut {NoStop}%
\bibitem [{\citenamefont {Joulain}\ \emph {et~al.}(2003)\citenamefont
  {Joulain}, \citenamefont {Carminati}, \citenamefont {Mulet},\ and\
  \citenamefont {Greffet}}]{Joulain:2003}%
  \BibitemOpen
  \bibfield  {author} {\bibinfo {author} {\bibfnamefont {K.}~\bibnamefont
  {Joulain}}, \bibinfo {author} {\bibfnamefont {R.}~\bibnamefont {Carminati}},
  \bibinfo {author} {\bibfnamefont {{\relax J.-P}.}~\bibnamefont {Mulet}}, \
  and\ \bibinfo {author} {\bibfnamefont {{\relax J.-J}.}~\bibnamefont
  {Greffet}},\ }\href {\doibase 10.1103/PhysRevB.68.245405} {\bibfield
  {journal} {\bibinfo  {journal} {Phys. Rev. B}\ }\textbf {\bibinfo {volume}
  {68}},\ \bibinfo {pages} {245405} (\bibinfo {year} {2003})}\BibitemShut
  {NoStop}%
\bibitem [{\citenamefont {Vielma}\ and\ \citenamefont
  {Leung}(2007)}]{Vielma:2007}%
  \BibitemOpen
  \bibfield  {author} {\bibinfo {author} {\bibfnamefont {J.}~\bibnamefont
  {Vielma}}\ and\ \bibinfo {author} {\bibfnamefont {P.}~\bibnamefont {Leung}},\
  }\href {\doibase 10.1063/1.2734549} {\bibfield  {journal} {\bibinfo
  {journal} {J. Chem. Phys.}\ }\textbf {\bibinfo {volume} {126}},\ \bibinfo
  {pages} {194704} (\bibinfo {year} {2007})}\BibitemShut {NoStop}%
\bibitem [{\citenamefont {Schimpf}\ \emph {et~al.}(2014)\citenamefont
  {Schimpf}, \citenamefont {Thakkar}, \citenamefont {Gunthardt}, \citenamefont
  {Masiello},\ and\ \citenamefont {Gamelin}}]{Schimpf:2014}%
  \BibitemOpen
  \bibfield  {author} {\bibinfo {author} {\bibfnamefont {{\relax
  A.M}.}~\bibnamefont {Schimpf}}, \bibinfo {author} {\bibfnamefont
  {N.}~\bibnamefont {Thakkar}}, \bibinfo {author} {\bibfnamefont {{\relax
  C.E}.}~\bibnamefont {Gunthardt}}, \bibinfo {author} {\bibfnamefont {{\relax
  D.J}.}~\bibnamefont {Masiello}}, \ and\ \bibinfo {author} {\bibfnamefont
  {{\relax D.R}.}~\bibnamefont {Gamelin}},\ }\href {\doibase 10.1021/nn406126u}
  {\bibfield  {journal} {\bibinfo  {journal} {ACS Nano}\ }\textbf {\bibinfo
  {volume} {8}},\ \bibinfo {pages} {1065} (\bibinfo {year} {2014})}\BibitemShut
  {NoStop}%
\bibitem [{\citenamefont {Zhang}\ \emph {et~al.}(2014)\citenamefont {Zhang},
  \citenamefont {Kulkarni}, \citenamefont {Prodan}, \citenamefont
  {Nordlander},\ and\ \citenamefont {Govorov}}]{ZhangNordlander:2014}%
  \BibitemOpen
  \bibfield  {author} {\bibinfo {author} {\bibfnamefont {H.}~\bibnamefont
  {Zhang}}, \bibinfo {author} {\bibfnamefont {V.}~\bibnamefont {Kulkarni}},
  \bibinfo {author} {\bibfnamefont {E.}~\bibnamefont {Prodan}}, \bibinfo
  {author} {\bibfnamefont {P.}~\bibnamefont {Nordlander}}, \ and\ \bibinfo
  {author} {\bibfnamefont {{\relax A.O}.}~\bibnamefont {Govorov}},\ }\href
  {\doibase 10.1021/jp5046035} {\bibfield  {journal} {\bibinfo  {journal} {J.
  Phys. Chem. C}\ }\textbf {\bibinfo {volume} {118}},\ \bibinfo {pages} {16035}
  (\bibinfo {year} {2014})}\BibitemShut {NoStop}%
\bibitem [{Note1()}]{Note1}%
  \BibitemOpen
  \bibinfo {note} {For completeness, we note that for a general multipole order
  $l$ the results of Eq.~\protect \textup {\hbox {\mathsurround \z@ \protect
  \normalfont (\ignorespaces \ref {eq:drudegraphene_intra}\unskip \@@italiccorr
  )}} generalize according to the substitutions $\omega _{\protect \mathrm
  {g}R}^2\rightarrow \protect \genfrac {}{}{}1{l+1}{2}\omega _{\protect \mathrm
  {g}R}^2$ and $\varepsilon _2\rightarrow \protect \genfrac
  {}{}{}1{l+1}{l}\varepsilon _2$.}\BibitemShut {Stop}%
\bibitem [{\citenamefont {Gong}\ \emph {et~al.}(2010)\citenamefont {Gong},
  \citenamefont {Lee}, \citenamefont {Shan}, \citenamefont {Vogel},
  \citenamefont {Wallace},\ and\ \citenamefont {Cho}}]{Gong:2010}%
  \BibitemOpen
  \bibfield  {author} {\bibinfo {author} {\bibfnamefont {C.}~\bibnamefont
  {Gong}}, \bibinfo {author} {\bibfnamefont {G.}~\bibnamefont {Lee}}, \bibinfo
  {author} {\bibfnamefont {B.}~\bibnamefont {Shan}}, \bibinfo {author}
  {\bibfnamefont {{\relax E.M}.}~\bibnamefont {Vogel}}, \bibinfo {author}
  {\bibfnamefont {{\relax R.M}.}~\bibnamefont {Wallace}}, \ and\ \bibinfo
  {author} {\bibfnamefont {K.}~\bibnamefont {Cho}},\ }\href {\doibase
  10.1063/1.3524232} {\bibfield  {journal} {\bibinfo  {journal} {J. Appl.
  Phys.}\ }\textbf {\bibinfo {volume} {108}},\ \bibinfo {pages} {123711}
  (\bibinfo {year} {2010})}\BibitemShut {NoStop}%
\bibitem [{\citenamefont {Kong}\ \emph {et~al.}(2010)\citenamefont {Kong},
  \citenamefont {Bjelkevig}, \citenamefont {Gaddam}, \citenamefont {Lee},
  \citenamefont {Han}, \citenamefont {Jeong}, \citenamefont {Wu}, \citenamefont
  {Zhang}, \citenamefont {Xiao}, \citenamefont {Dowben},\ and\ \citenamefont
  {Kelber}}]{Kong:2010}%
  \BibitemOpen
  \bibfield  {author} {\bibinfo {author} {\bibfnamefont {L.}~\bibnamefont
  {Kong}}, \bibinfo {author} {\bibfnamefont {C.}~\bibnamefont {Bjelkevig}},
  \bibinfo {author} {\bibfnamefont {S.}~\bibnamefont {Gaddam}}, \bibinfo
  {author} {\bibfnamefont {{\relax Y.H}.}~\bibnamefont {Lee}}, \bibinfo
  {author} {\bibfnamefont {{\relax G.H}.}~\bibnamefont {Han}}, \bibinfo
  {author} {\bibfnamefont {{\relax H.K}.}~\bibnamefont {Jeong}}, \bibinfo
  {author} {\bibfnamefont {N.}~\bibnamefont {Wu}}, \bibinfo {author}
  {\bibfnamefont {Z.}~\bibnamefont {Zhang}}, \bibinfo {author} {\bibfnamefont
  {J.}~\bibnamefont {Xiao}}, \bibinfo {author} {\bibfnamefont {{\relax
  P.A}.}~\bibnamefont {Dowben}}, \ and\ \bibinfo {author} {\bibfnamefont
  {{\relax J.A}.}~\bibnamefont {Kelber}},\ }\href {\doibase 10.1021/jp108616h}
  {\bibfield  {journal} {\bibinfo  {journal} {J. Phys. Chem. C}\ }\textbf
  {\bibinfo {volume} {114}},\ \bibinfo {pages} {21618} (\bibinfo {year}
  {2010})}\BibitemShut {NoStop}%
\bibitem [{\citenamefont {Kreibig}\ and\ \citenamefont
  {Genzel}(1985)}]{Kreibig:1985}%
  \BibitemOpen
  \bibfield  {author} {\bibinfo {author} {\bibfnamefont {U.}~\bibnamefont
  {Kreibig}}\ and\ \bibinfo {author} {\bibfnamefont {L.}~\bibnamefont
  {Genzel}},\ }\href {\doibase 10.1016/0039-6028(85)90239-0} {\bibfield
  {journal} {\bibinfo  {journal} {Surf. Sci.}\ }\textbf {\bibinfo {volume}
  {156}},\ \bibinfo {pages} {678} (\bibinfo {year} {1985})}\BibitemShut
  {NoStop}%
\bibitem [{\citenamefont {Apell}\ and\ \citenamefont
  {Penn}(1983)}]{Apell:1983}%
  \BibitemOpen
  \bibfield  {author} {\bibinfo {author} {\bibfnamefont {P.}~\bibnamefont
  {Apell}}\ and\ \bibinfo {author} {\bibfnamefont {{\relax D.R}.}~\bibnamefont
  {Penn}},\ }\href {\doibase 10.1103/PhysRevLett.50.1316} {\bibfield  {journal}
  {\bibinfo  {journal} {Phys. Rev. Lett.}\ }\textbf {\bibinfo {volume} {50}},\
  \bibinfo {pages} {1316} (\bibinfo {year} {1983})}\BibitemShut {NoStop}%
\bibitem [{\citenamefont {Christensen}\ \emph
  {et~al.}(2014{\natexlab{b}})\citenamefont {Christensen}, \citenamefont
  {Wang}, \citenamefont {Jauho}, \citenamefont {Wubs},\ and\ \citenamefont
  {Mortensen}}]{Christensen:2014prb}%
  \BibitemOpen
  \bibfield  {author} {\bibinfo {author} {\bibfnamefont {T.}~\bibnamefont
  {Christensen}}, \bibinfo {author} {\bibfnamefont {W.}~\bibnamefont {Wang}},
  \bibinfo {author} {\bibfnamefont {{\relax A.-P}.}~\bibnamefont {Jauho}},
  \bibinfo {author} {\bibfnamefont {M.}~\bibnamefont {Wubs}}, \ and\ \bibinfo
  {author} {\bibfnamefont {{\relax N.A}.}~\bibnamefont {Mortensen}},\ }\href
  {10.1103/PhysRevB.90.241414} {\bibfield  {journal} {\bibinfo  {journal}
  {Phys. Rev. B}\ }\textbf {\bibinfo {volume} {90}},\ \bibinfo {pages}
  {241414(R)} (\bibinfo {year} {2014}{\natexlab{b}})}\BibitemShut {NoStop}%
\bibitem [{\citenamefont {Jackson}(1999)}]{Jackson}%
  \BibitemOpen
  \bibfield  {author} {\bibinfo {author} {\bibfnamefont {{\relax
  J.D}.}~\bibnamefont {Jackson}},\ }\href
  {http://eu.wiley.com/WileyCDA/WileyTitle/productCd-047130932X.html} {\emph
  {\bibinfo {title} {{C}lassical {E}lectrodynamics}}},\ \bibinfo {edition}
  {3rd}\ ed.\ (\bibinfo  {publisher} {John Wiley \& Sons},\ \bibinfo {address}
  {New York},\ \bibinfo {year} {1999})\BibitemShut {NoStop}%
\end{thebibliography}
\end{document}